\documentclass[10pt]{iopart}
\usepackage{fullpage}
\usepackage{xcolor}
\usepackage[latin1]{inputenc}
\usepackage{amsfonts}
\usepackage{amssymb}
\expandafter\let\csname equation*\endcsname\relax
\expandafter\let\csname endequation*\endcsname\relax
\usepackage{amsmath}
\usepackage[colorlinks]{hyperref}
\usepackage{rotating,graphicx,tikz,scalerel}
\usepackage[numbers,sort&compress]{natbib} 
\usepackage[british]{babel}

\usepackage{caption}
\usepackage{subcaption}

\usetikzlibrary{svg.path}
\definecolor{orcidlogocol}{HTML}{A6CE39}
\tikzset{
  orcidlogo/.pic={
    \fill[orcidlogocol] svg{M256,128c0,70.7-57.3,128-128,128C57.3,256,0,198.7,0,128C0,57.3,57.3,0,128,0C198.7,0,256,57.3,256,128z};
    \fill[white] svg{M86.3,186.2H70.9V79.1h15.4v48.4V186.2z}
                 svg{M108.9,79.1h41.6c39.6,0,57,28.3,57,53.6c0,27.5-21.5,53.6-56.8,53.6h-41.8V79.1z M124.3,172.4h24.5c34.9,0,42.9-26.5,42.9-39.7c0-21.5-13.7-39.7-43.7-39.7h-23.7V172.4z}
                 svg{M88.7,56.8c0,5.5-4.5,10.1-10.1,10.1c-5.6,0-10.1-4.6-10.1-10.1c0-5.6,4.5-10.1,10.1-10.1C84.2,46.7,88.7,51.3,88.7,56.8z};
  }
}
\newcommand\orcidicon[1]{\href{https://orcid.org/#1}{\mbox{\scalerel*{
\begin{tikzpicture}[yscale=-1,transform shape] \pic{orcidlogo}; \end{tikzpicture}
}{|}}}}

\graphicspath{{pics/}}
\DeclareMathAlphabet{\mathpzc}{OT1}{pzc}{m}{it}
\hypersetup{colorlinks,    citecolor=blue,    filecolor=black,   linkcolor=blue,  urlcolor=blue}

\newcommand{\newblock}{}
\renewcommand{\theenumi}{\roman{enumi}}%

\begin{document}
\title{Electron impact single ionization of hydrogen molecule by twisted electron beam}
\author{{Nikita Dhankhar \orcidicon{0000-0001-9423-4796}}}

\author{{R. Choubisa \orcidicon{0000-0003-3000-6174}} }
\ead{rchoubisa@pilani.bits-pilani.ac.in}
\address{Department of Physics, Birla Institute of Technology and Science-Pilani, Pilani Campus, Pilani,  Rajasthan,  333031, India}

\begin{abstract}
In this communication, we present the results of the Five-fold Differential Cross Section (5DCS) and  Triple Differential Cross Section (TDCS) for the (e,2e) process on molecular hydrogen ($H_2$) by the plane wave and the twisted electron beam impact. The formalism is developed within the first Born approximation using the plane wave and the twisted wave for the incident electron beam. We describe the plane wave, Heitler-London type wave function and Coulomb wave for the scattered electron, $H_2$ molecular state, and the ejected electron respectively. We compare the angular profiles of the 5DCS and TDCS for the different values of Orbital Angular Momentum (OAM) number {\it m} of the twisted electron beam with that of the plane wave beam. We also present the 5DCS for different molecular orientations and study the effect of {\it m} on the 5DCS. We further investigate the influence of the twisted electron beam on the (e,2e) process on the $H_2$ molecule from the perspective of the ``Young-type" interference of the scattered waves, emanating from the two atomic centers of the $H_2$ molecule. We also study the TDCS for macroscopic $H_2$ target to explore the effect of opening angle ($\theta_p$) of the twisted electron beam on the TDCS. Our results clearly show the effect of the twisted electron's OAM number (m) and the opening angle ($\theta_p$) on the 5DCS and TDCS of the molecular hydrogen.

\end{abstract}

\maketitle
\ioptwocol

\section{Introduction}
Electron interactions with the atomic and molecular targets play an important role in biology, astronomy, plasma physics, radiation physics
and chemistry \cite{Bart2016, Shalenov2017, Bug2017, De2019, Chavez2019}. The electron impact single ionization process, in which we determine the momenta of the two outgoing electrons, is known as an (e,2e) process. Hence in an (e,2e) process, the two outgoing electrons are detected in coincidence with their energies and directions fully resolved \cite{Camp2018}. The first experiment in the field of (e,2e) was performed independently by Amaldi \textit{et al.} \cite{Amaldi1969} and Ehrhardt \textit{et al.} \cite{Ehrhardt1969}. Since then, many coincident experiments were performed on the atomic and molecular targets in various geometrical arrangements of the detected electrons. These experiments further led to the development of numerous theoretical models of the (e,2e) reactions on atomic and molecular targets. (e,2e) studies help us to understand the structure of targets, electron-electron correlations, and reaction mechanisms during the ionization processes.

Tremendous progress has been made in the field of electron impact ionization studies for atomic targets \cite{Ren2015, Whelan2012, Lahmam1991, Casagrande2008, Amami2017},  both at the experimental and theoretical level. The (e,2e) study was further extended to molecular targets to investigate the effects of multi-centers on (e,2e) angular distributions of the Triple Differential Cross Section (TDCS). The electron impact ionization study for molecular targets is different from that for atoms because of the molecular configuration of the targets arising from the different molecular orbitals involved \cite{Colyer2009}. In literature, numerous studies in the (e,2e) field have been done on various molecules, like $H_2$, $N_2$, $O_2$, $H_2O$ $CO_2$, formic acid, bio-molecules such as DNA, RNA etc. \cite{Colgan2009,Mouawad2017,Ren2017,Li2017,Hossen2018,Sakaamini2018,Khatir2019,Singh2019,Ali2020}. 

	The simplest molecule, hydrogen molecule ($H_2$), has drawn lots of attention for the study of (e,2e) processes on molecules. In addition to (e,2e) processes , various theoretical and experimental investigations have been carried out to study the double ionization (\textit{i.e.} (e,3e) process) on $H_2$ \cite{Mansouri2004,Chul2011,Pindzola2018}. The molecular hydrogen is a basic prototype molecule for the study of the ``Young-type" interference effect caused by the scattering of the incident particle by the two centers of the $H_2$. Since the two atoms in the $H_2$ molecule are indistinguishable, we can envisage the $H_2$ molecule as two sources of secondary matter-wave emitters, which will lead to the interference effect. The Young's type interference effect was theoretically predicted by photon impact on $H_2$ by Cohen and Fano (1966) \cite{Cohen1966}. The interference effect was also observed for electron emission on $H_2$ by fast ion impact \cite{Stolterfoht2001, Stolterfoht2003}. Different groups have studied this aspect, both at the theoretical and experimental level, by studying the ejected electron's angular distribution \cite{Brownlie2006, Staicu2008, Fojon2006, Stia2003,Ciappina2014,Li2018}. To the best of our knowledge, almost all of the (e,2e) collision studies on molecular targets have been done for the plane wave electron beam (not possessing any orbital angular momentum).

A vortex beam characterizes a freely propagating beam possessing a helical wave-front with a well-defined Orbital Angular Momentum (OAM), {\it m}, along the propagation direction. The theoretical study by Bliokh \textit{et al.} \cite{Bliokh2007} for non-relativistic electrons advanced the experimental and theoretical investigations of the electron states with vortices  (i.e., Electron Vortex Beams (EVBs), twisted electron beams). The experimental observation of the electron vortex states accentuated an exciting area of collision studies for the study of their interaction with the atoms and molecules \cite{Uchida2010, Verbeeck2010, Morran2011}. Unlike the plane waves, twisted electron beams carry a non-zero OAM ({\it m}) projection along the propagation direction. These beams have a helical phase front $e^{im\phi}$\ with the azimuthal angle $\phi$ about the propagation axis (see \cite{Llyod2017, Bliokh2017, Hugo2018} for a better theoretical insight of EVBs).
Twisted electron beams provide scope for research in optical microscopy, quantum state manipulation, optical tweezers, astronomy, higher-order harmonic generation and many more fields \cite{Neil2002, Spiral2005, Berkhout2009, Verbeeck2010, Morran2011,Sebastian2019}.

 It is essential to understand how electrons with non-zero OAM interact with atoms for applications of the twisted electron collisions with atoms. A handful of studies are available for ionization/scattering of atomic targets using twisted beams. The work by Ivanov and Serbo (2011) \cite{Serbo2011} and Boxem and co-workers (2014) \cite{Boxem2014} contributed to the outset of the theoretical analysis of the scattering experiments by twisted electrons.  Boxem \textit{et al.} \cite{Boxem2015} studied the inelastic scattering amplitudes analytically for the hydrogen-like systems determining the influence of the OAM on the scattering amplitudes. Serbo \textit{et al.} \cite{Serbo2015} analyzed the scattering by twisted electrons in the relativistic framework illustrating that the angular distribution and the polarization of the outgoing electrons depend not only on the transverse and the longitudinal momenta but also on the projection of the Total Angular Momentum (TAM) along the propagation direction of the beam. The collisional studies for hydrogenic atom, treating EVBs as spatially localized wave packets, by Karlovets \textit{et al.} \cite{Karlovets2017} showed that the number of scattering events in collisions involving twisted electrons is comparable to that in the standard plane-wave calculations. Maiorova \textit{at al.} \cite{Maiorova2018} advanced the scattering studies by their theoretical analysis of the Differential Cross Section (DCS) for the molecular hydrogen $H_2$. The DCS emphasized the influence of the twisted electron on the Young-type interference. Harris \textit{et al.} (2019) reported the ionization of the Hydrogen atom by twisted electrons by analyzing the fully differential cross section (FDCS) with different parameters of the twisted electron beam \cite{Harris2019}. The results indicated a shift in the binary and recoil peak for twisted electron's cross-sections from their plane wave locations due to the projectile's transverse momentum components. The recent study by Mandal \textit{et al.} (2020) showed the dependence of the TAM number (\textit{m}) on the angular profile of the TDCS and spin asymmetry for the relativistic electron impact ionization of the heavy atomic targets \cite{Mandal2020}. Furthermore, Dhankhar \textit{et al.} \cite{Dhankhar2020}  studied theoretically the double ionization of He atom in $\theta$-variable and constant $\theta_{12}$ mode for the twisted electron incidence. Their results found that the angular profile of the Five-Fold Differential Cross-section (FDCS) depends on the OAM number \textit{m} and the opening angle $\theta_p$ of the incident twisted electron beam.

To the best of our knowledge, the cross-section studies for the (e,2e) processes by the twisted electron beams have been performed mostly for the atoms. In this communication, we present the first theoretical estimation for the (e,2e) process of the $H_2$ molecule for the twisted electron case. We develop our formalism within the first Born approximation framework for the plane wave electron beam (Sec.\ \ref{sec2a}) and the twisted electron beam in (Sec.\ \ref{sec2b}). We describe the plane wave, Heitler-London type wave function, and Coulomb wave for the scattered electron, $H_2$ molecular state, and the ejected electron, respectively. We present the Five-fold Differential Cross-Section(5DCS) and Triple Differential Cross Section (TDCS) in the coplanar asymmetric geometry for different parameters of the twisted electron beam in Sec.\ \ref{sec3}. Finally, we conclude our paper in the Sec.\ \ref{sec4}. Atomic units are used throughout the paper unless otherwise stated.

\section{Theory}\label{sec2}
In this section, we present the theoretical formalism for the (e,2e) process of $H_2$ molecule by considering both the plane wave and the twisted electron beam as an incident beam.

\subsection{Plane-wave (e,2e) ionization}\label{sec2a}
\begin{figure}[ht]
\includegraphics[width = 0.9\columnwidth]{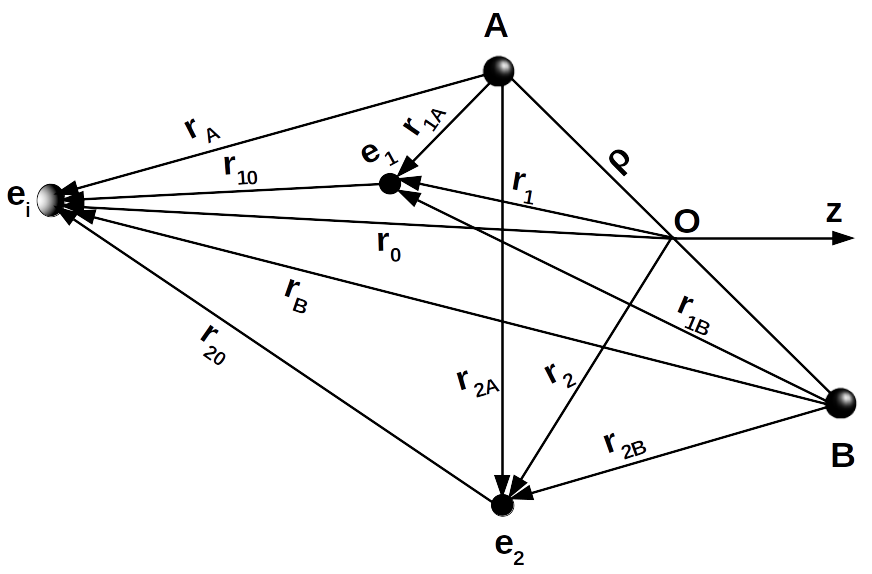}
\caption{Coordinates representation of the hydrogen molecule. $\mathbf{e}_i$ represents the incident electron with energy $E_i$. The molecule is centered at origin O which is located at the center of the inter-nuclear vector $\boldsymbol{\rho}$. A and B represents the two center of the $H_2$ molecule. $\mathbf{e}_j$({\it j}=1,2) represents the first and second electron from the center of the $H_2$ molecule. $\mathbf{r}_0$ is the radial vector of the incident electron. $\mathbf{r}_{j0}$({\it j}=1,2) represents the position vector of the $j^{th}$ electron from the incident electron. $\mathbf{r}_j$({\it j}=1,2) is the position vector of $j^{th}$ electron from the origin O. $\mathbf{r}_{jA}$({\it j}=1,2) and $\mathbf{r}_{jB}$({\it j}=1,2) are coordinates of the $j^{th}$ electron from the atomic centers A and B respectively.}\label{fig1}
\end{figure}

In an (e,2e) process on $H_2$ molecule,  an incident electron with energy $E_i$ and momentum $\mathbf{k}_i$ collides with the target resulting in ejection of bound electron with energy $E_e$ with momentum $\mathbf{k}_e$. The ejected electron and the scattered electron with energy $E_s$ and momentum $\mathbf{k}_s$ are detected and measured coincidentally. The electron impact ionization of $H_2$ can be standardized as; 
\begin{equation}\label{e1}
e_i(\mathbf{k}_i) \ +\ H_2 \rightarrow\ H_2^+ \ +\ e_s(\mathbf{k}_s)\ +\ e_e(\mathbf{k}_e).
\end{equation}
The Five-fold Differential Cross Section (5DCS) describes the complete picture of the (e,2e) process on $H_2$. In the first Born approximation for an (e,2e) process, 5DCS can be described as; 
\begin{equation}\label{e2}
\frac{d^5\sigma}{dE_ed\Omega_sd\Omega_ed\Omega_{\rho}} = (2\pi)^4\ \frac{k_ek_s}{k_i}\ |T_{fi}|^2 ,
\end{equation}
here $dE_e$ describes the energy interval for the ejected electron and $d\Omega_s$, $d\Omega_e$ and $d\Omega_{\rho}$ are the solid angle's intervals of the scattered, ejected electron and the inter-nuclear vector $\boldsymbol{\rho}$  respectively. The scattering amplitude is given by,
\begin{equation}\label{e3}
T_{fi} =\ \langle \phi_f|V|\phi_i \rangle,
\end{equation}
where $\phi_i$ is the initial state wave function, consists of the incident electron and $H_2$ target, $\phi_f$ is the final state wave-function consists of the continuum state of the outgoing electrons and the remaining target and V is the interaction potential between the incident electron and $H_2$ molecule. For an (e,2e) process, in the FBA, $T_{fi}$ is expressed as;
\begin{equation}\label{e4}
T_{fi} = \ \langle \varphi_{\mathbf{k}_s,\mathbf{k}_e}(\boldsymbol{\rho},\mathbf{r}_0,\mathbf{r}_1,\mathbf{r}_2)|V|\varphi_{\mathbf{k}_i}(\boldsymbol{\rho},\mathbf{r}_0,\mathbf{r}_1,\mathbf{r}_2)\rangle,
\end{equation}
where $\varphi_{\mathbf{k}_s,\mathbf{k}_e}(\boldsymbol{\rho},\mathbf{r}_0,\mathbf{r}_1,\mathbf{r}_2)$ is the continuum state of the scattered and the ejected electrons in the presence of $H^+$ ion. $\varphi_{\mathbf{k}_i}(\boldsymbol{\rho},\mathbf{r}_0,\mathbf{r}_1,\mathbf{r}_2)$ is the initial state consisting of the incident electron and the molecular bound state of the $H_2$ molecule. $\mathbf{r}_j$({\it j}=1,2) represents the radial vector the $j^{th}$ electron with respect to the center of the molecule, assuming that the center of the molecule is at the origin (see figure \ref{fig1}). In figure \ref{fig1}, $\mathbf{r}_0$ is the radial vector of the incident electron and $\boldsymbol{\rho}$ is the equilibrium internuclear vector of the molecule. The interaction potential V can be represented as,
\begin{equation}\label{e5}
V = \frac{1}{|\mathbf{r}_0 - \mathbf{r}_1|}+\frac{1}{|\mathbf{r}_0 - \mathbf{r}_2|}-\frac{1}{|\mathbf{r}_0 - \frac{\boldsymbol{\rho}}{2}|}-\frac{1}{|\mathbf{r}_0 + \frac{\boldsymbol{\rho}}{2}|}.\\
\end{equation}
Here, we use the plane wave for the incident and the scattered electrons. The normalized plane wave for the incident electrons can be expressed as;
\begin{equation}\label{e6}
\phi_i(\mathbf{r}_0) = \frac{1}{(2\pi)^{3/2}}e^{i\mathbf{k}_i \cdot \mathbf{r}_0},
\end{equation}
with the incident electron momentum $\mathbf{k}_i$.
The target state wave function, characterizing the molecular bound sate of $H_2$, is described by the Heitler-London type wave function \cite{Wang1928},
\begin{equation}\label{e7}
\psi_i(\boldsymbol{\rho},\mathbf{r}_1,\mathbf{r}_2) = N_{HL}(\rho)\{e^{-\alpha r_{1A}}e^{-\alpha r_{2B}} + e^{-\alpha r_{2A}}e^{-\alpha r_{1B}}\},
\end{equation}
with $\alpha$ = 1.166, $\rho$ = 1.406 and $N_{HL}$ is the normalization constant. Here, $r_{jA}$({\it j} = 1,2) and  $r_{jB}$({\it j} = 1,2) are defined in figure \ref{fig1}.

The final state wave function is given as \cite{Stia2002},
\begin{equation}\label{e8}
\psi_f^-(\boldsymbol{\rho},\mathbf{r}_1,\mathbf{r}_2) = \phi_f(\rho,\mathbf{r}_2)\phi_c,
\end{equation}
where $\phi_f$ is the bound state of $H_2^+$ and $\phi_c$ is the Coulomb wave function.
$\phi_f$ is represented as the linear combination of atomic orbitals,
\begin{equation}\label{e9}
\phi_f(\rho,\mathbf{r}_2) = N_f(\rho)\{ e^{-\beta r_{2A}}+e^{-\beta r_{2B}} \},
\end{equation}
and the Coulomb wave function $\phi_c$ is given by,
\begin{equation}\label{e10}
\phi_c = e^{i\mathbf{k}_e \cdot \mathbf{r}_1}C _1F_1(i\gamma;1;-i(k_er_{1j}+\mathbf{k}_e \cdot \mathbf{r}_{1j})),
\end{equation}
where j = A or B. The function $\phi_c$ correlate with j = A(or B) when the exponential term in equation (\ref{e7}) is acknowledged \cite{Stia2002}. Also, C = $(2\pi)^{-3/2} \Gamma(1-i\gamma)e^{-\pi/2}$ where $\Gamma(1-i\gamma)$ is the Gamma function, $\gamma = -1/k_e$ and $_1F_1(i\gamma;1;-i(kr+\mathbf{k}_e \cdot \mathbf{r}_{1j}))$ is the confluent hyper-geometric function. 
Using the above equations, the transition matrix, $T_{fi}$, can be written as;
\begin{equation}\label{e11}
\begin{split}
T_{fi} = \frac{1}{(2\pi)^{9/2}}N_{HL}N_f\Gamma(1+i\gamma)exp(-\pi\gamma/2) \times \\ 
\Big\langle e^{i\mathbf{k}_e \cdot \mathbf{r}_1} _1F_1(i\gamma;1;-i(k_er_{1j}+\mathbf{k}_e\cdot\mathbf{r}_{1j})(e^{-\beta r_{2A}}+e^{-\beta r_{2B}})\\
\times \ |\ V\ | \ e^{i\mathbf{K} \cdot \mathbf{r}_0}(e^{-\alpha r_{1A}}e^{-\alpha r_{2B}} + e^{-\alpha r_{2A}}e^{-\alpha r_{1B}})\Big\rangle,
\end{split}
\end{equation}
where $\mathbf{K} = \mathbf{k}_i-\mathbf{k}_s$ is the momentum transferred to the $H_2$ molecule. Following \cite{Stia2002}, the transition matrix element computation in equation (\ref{e11}) can be interpreted as contribution from direct terms and indirect terms. The direct terms may be attributed to the ionization of e1 from the center j = A(or B) due to interaction of incident electron with e1 and center j. While, the indirect term recognizes the ionization of  e1 from center j via interaction of the incident electron with e2 and center B. Due to the large difference in the energies of the ejected electrons with that of the incident and scattered electrons ($E_i, E_s \gg E_e$), the contribution of the indirect terms in the transition matrix is negligible. The transition matrix element can then be described as;
\begin{equation}\label{e12}
\begin{split}
T_{fi} = \frac{1}{(2\pi)^{9/2}}N_{HL}N_f\Gamma(1+i\gamma)exp(-\pi\gamma/2) \times \\   \sum_{\substack{j,k\\ j\neq k}} \Big\langle e^{i\mathbf{k}_e \cdot \mathbf{r}_1} _1F_1(i\gamma;1;-i(k_er_{1j}+\mathbf{k}_e \cdot \mathbf{r}_{1j})(e^{-\beta r_{2j}}+e^{-\beta r_{2k}})\\
\times \ \Big|\ \Big( \frac{1}{\mathbf{r}_{10}}-\frac{1}{\mathbf{r}_j} \Big) \ \Big| \ e^{i\mathbf{K} \cdot \mathbf{r}_0}(e^{-\alpha r_{1j}}e^{-\alpha r_{2k}})\Big\rangle.
\end{split}
\end{equation}
The integration over $\mathbf{r}_0$ can be done analytically. Following the results from Tweed \cite{Tweed1992} we have,
\begin{equation}\label{e13}
\int \frac{e^{i\mathbf{K} \cdot \mathbf{r}_0}}{|\mathbf{r}_0 - \mathbf{r}_1|} d^3r_0 = \frac{4\pi}{K^2}e^{i\mathbf{K} \cdot \mathbf{r}_1}. 
\end{equation}
The matrix element $T_{fi}$ is thus given by,
\begin{equation}\label{e14}
T_{fi}(\mathbf{K}) = \langle \psi^-_{f} | 2 e^{i\mathbf{K} \cdot \mathbf{r}_1} - e^{-i\mathbf{K} \cdot \boldsymbol{\rho}/2} - e^{i\mathbf{K} \cdot \boldsymbol{\rho}/2}| \psi_i \rangle.
\end{equation}
The transition matrix element given by equation (\ref{e14}) can be written in a convenient form, such as;
\begin{equation}\label{e15}
|T_{fi}|^2 \cong 2 |[1+\cos(\mathbf{q} \cdot \boldsymbol{\rho})]T_{fi}^A|^2,
\end{equation}
where $\mathbf{q} =  \mathbf{K}-\mathbf{k}_e$ is the momentum transferred to the residual ion and $T_{fi}^A$ is the transition matrix element corresponding to the (e,2e) process for the one-centred atomic hydrogen. Since all the directions of the inter-nuclear vector $\boldsymbol{\rho}$ are equally probable for the $H_2$ molecule, the integration over all the orientations of $\boldsymbol{\rho}$ gives the proper orientation average TDCS which can be defined as;
\begin{equation}\label{e16}
 \frac{d^3\sigma}{dE_e d\Omega_s d\Omega_e} =  \frac{1}{4\pi}\int_{\phi_{\rho}=0}^{2\pi} \int_{\theta_{\rho}=0}^{\pi} \frac{d^5\sigma}{dE_ed\Omega_sd\Omega_ed\Omega_{\rho}} \ d\Omega_{\rho} 
\end{equation} 
where $\frac{d^5\sigma}{dE_ed\Omega_sd\Omega_ed\Omega_{\rho}}$ is the orientation dependent 5DCS given by equation (\ref{e2}) and $\theta_{\rho}$, $\phi_{\rho}$  are the standard polar and azimuthal angles of $\boldsymbol{\rho}$ respectively assuming that the propagation of the electron beam is along the {\it z}-axis. 

\subsection{Twisted electron (e,2e) ionization}\label{sec2b}
\begin{figure}[ht]
\includegraphics[width = 1\columnwidth]{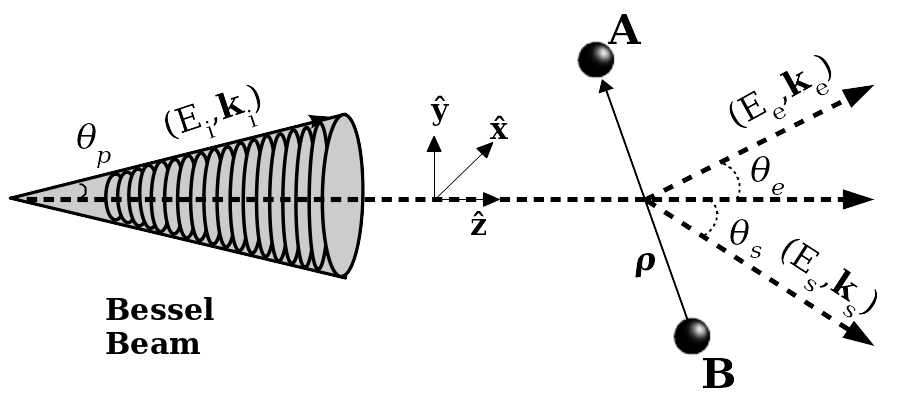}
\caption{Schematic diagram for the electron impact ionization of $H_2$ molecule by Bessel beam. $\theta_p$ is the opening angle of the beam. An incident twisted electron of energy $E_i$ and momentum $\mathbf{k}_i$ interacts with the $H_2$ molecule ejecting one of the bound electrons of the molecule into continuum state. We describe $E_s$ and $E_e$ and $\mathbf{k}_s$ and $\mathbf{k}_e$ as the energies and momenta of the scattered and ejected electron respectively.  $\theta_s$ and $\theta_e$ represent the scattered and ejected electrons' angular positions. $\Omega_{\rho}$ corresponds to the solid angle of the internuclear vector $\boldsymbol{\rho}$. The quantization (\textit{z})-axis is chosen along the propagation direction of the incoming beam. The beam propagates out of the page and twists around the propagation direction (counter-clockwise).} \label{fig2}
\end{figure}
Figure \ref{fig2} shows the (e,2e) process on $H_2$ molecule by the twisted electron beam. The twisted beam has a helical wave-front that twists around the beam axis when it propagates along the beam direction.
The formalism for the computation of the (e,2e) cross sections for the $H_2$ molecule is the same as used in Sec.\ref{sec2a}, except we replace the plane wave for the incident electron beam with a twisted electron beam, such as a Bessel beam.

The incident momentum vector $\mathbf{k}_i$ can be described as, 
\begin{equation}\label{e17}
\mathbf{k}_i = (k_i \sin\theta_p \cos\phi_p)\hat{x} + (k_i \sin\theta_p \sin\phi_p)\hat{y} + (k_i \cos\theta_p)\hat{z}
\end{equation}
with $\theta_p$  and $\phi_p$ defined as the polar and azimuthal angles of the $\mathbf{k}_i$. 
The longitudinal momentum is along the \textit{z}-axis, with the center of the $H_2$ molecule on the axis. $\mathbf{k}_i$ makes an angle $\theta_p$ with the \textit{z}-axis which is normally referred as the opening angle of the twisted beam. $\theta_p = \tan^{-1}\frac{k_{i\perp}}{k_{iz}}$ with  $k_{i\perp}$ and $k_{iz}$ are the perpendicular and the longitudinal components of the momentum  $\mathbf{k}_i$ respectively.

The initial state, $\varphi_i(\mathbf{r}_0,\mathbf{r}_1,\mathbf{r}_2)$, is the product of the Bessel beam wave function (twisted electron wave function), $\psi^{(tw)}_{\varkappa m}(\mathbf{r}_0)$ , and the target state wave function of $H_2$, $\psi_i(\boldsymbol{\rho},\mathbf{r}_1,\mathbf{r}_2)$, given by equation (\ref{e7}).

\begin{equation}\label{e18}
\varphi_i(\mathbf{r}_0,\mathbf{r}_1,\mathbf{r}_2) =
\psi^{(tw)}_{\varkappa m}(\mathbf{r}_0)\psi_i(\boldsymbol{\rho},\mathbf{r}_1,\mathbf{r}_2).
\end{equation}
 
As specified in \cite{Boxem2015}, the Bessel beam can be interpreted as superposition of plane waves such that
\begin{equation}\label{e19}
\psi^{(tw)}_{\varkappa m}(\mathbf{r}_0) = \int^{\infty}_{0} \frac{dk_{i\perp}}{2\pi}\ k_{i\perp} \int_{0}^{2\pi}\frac{d\phi_p}{2\pi}\ a_{\varkappa m}(\mathbf{k}_{i\perp})e^{i\mathbf{k}_i \cdot \mathbf{r}_0},
\end{equation}

with the amplitude
\begin{equation}\label{e20}
a_{\varkappa m}(\mathbf{k}_{i\perp}) = (-i)^m \ \sqrt{\frac{2\pi}{\varkappa}}\ e^{im\phi_p} \ \delta(|\mathbf{k}_{i\perp}| - \varkappa),
\end{equation}

where $\varkappa$ is the absolute value of the transverse momentum ($k_i \sin\theta_p$). The integration over the transverse incident momentum $k_{i\perp}$ results $\psi^{(tw)}_{\varkappa m}(\mathbf{r}_0)$ as;
\begin{equation}\label{e21}
\psi^{(tw)}_{\varkappa m}(\mathbf{r}_0) = (-i)^m \sqrt{\frac{\varkappa}{2\pi}} \int_{0}^{2\pi}\ \frac{d\phi_p}{2\pi}\ e^{im\phi}\ e^{i\mathbf{k}_i \cdot \mathbf{r}_0},
\end{equation}

The final state wave function is the product of the scattered plane wave and the final state wave function given by equation (\ref{e9}),
\begin{equation}\label{e22}
\varphi_{k_i,k_s,k_e}(\mathbf{r}_0,\mathbf{r}_1,\mathbf{r}_2) = \phi_{k_s}(\mathbf{r}_0)\psi_{f}^-(\boldsymbol{\rho},\mathbf{r}_1,\mathbf{r}_2),
\end{equation}
where $\phi_{k_s}(\mathbf{r}_0)$ is a plane wave given by 
\begin{equation}\label{e23}
 \phi_{k_s}(\mathbf{r}_0) = \frac{e^{i\mathbf{k}_s \cdot \mathbf{r}_0}}{(2\pi)^{3/2}}.
\end{equation}

Substituting equations (\ref{e18}) and (\ref{e22}) in equation (\ref{e4}), we can write the twisted wave transition amplitude ($T^{tw}_{fi}(\varkappa,\mathbf{K})$) in terms of the plane wave transition amplitude ($T_{fi}(\mathbf{K})$ given by equation (\ref{e14})) as, 
\begin{equation}\label{e24}
T^{tw}_{fi}(\varkappa,\mathbf{K}) = (-i)^m \sqrt{\frac{\varkappa}{2\pi}} \int \frac{d\phi_p}{2\pi}\ e^{im\phi_p}\ T_{fi}(\mathbf{K}).
\end{equation}
 
Here, the momentum transfer to the target can be expressed as; 
\begin{equation}\label{e25}
K^2 = k_i^2 + k_s^2 - 2k_ik_s \cos\theta ,
\end{equation}
where, 
\begin{equation}\label{e26}
\cos\theta = \cos\theta_p \cos\theta_s + \sin\theta_p \sin\theta_s \cos(\phi_p - \phi_s).
\end{equation}
Here, $\theta_s$ and $\phi_s$ are the polar and azimuthal angles of the $\mathbf{k}_s$. For the coplanar geometry $\phi_s = 0$. The TDCS for the twisted electron can be computed from equation \ref{e2} with $T_{fi}^{tw}(\varkappa, \mathbf{K})$ computed from the equation \ref{e24}.

\begin{figure*}
\centering
\begin{subfigure}{.5\textwidth}
  \centering
  \includegraphics[width=1.1\linewidth]{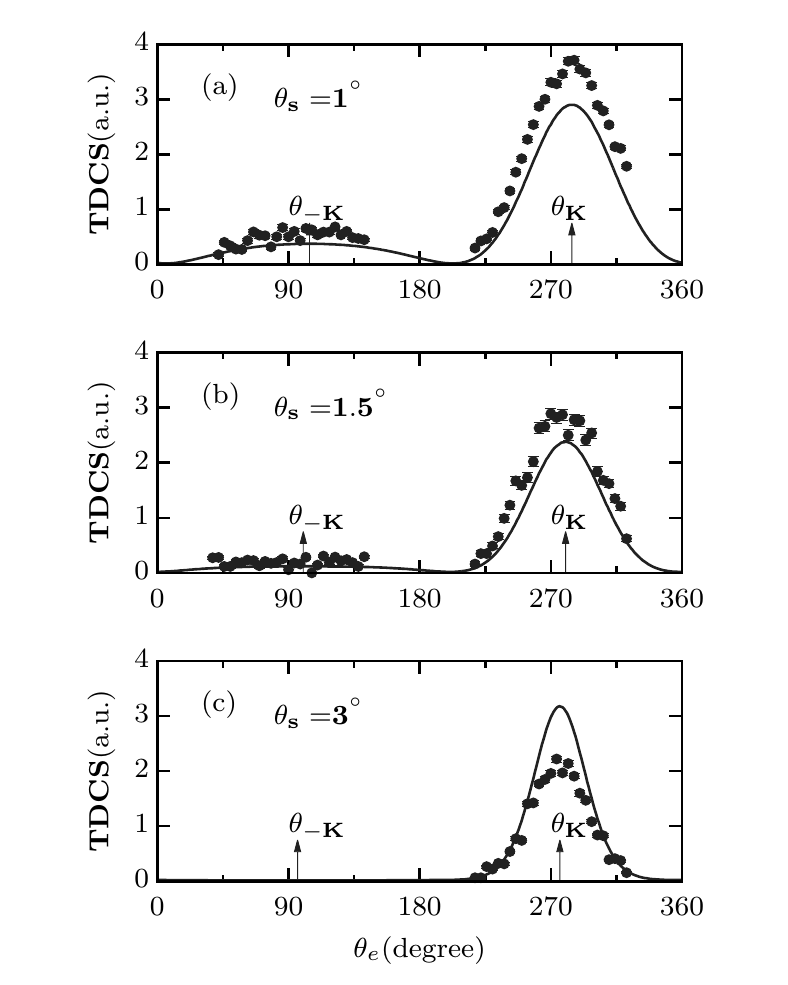}
\end{subfigure}%
\begin{subfigure}{.5\textwidth}
  \centering
  \includegraphics[width=1.1\linewidth]{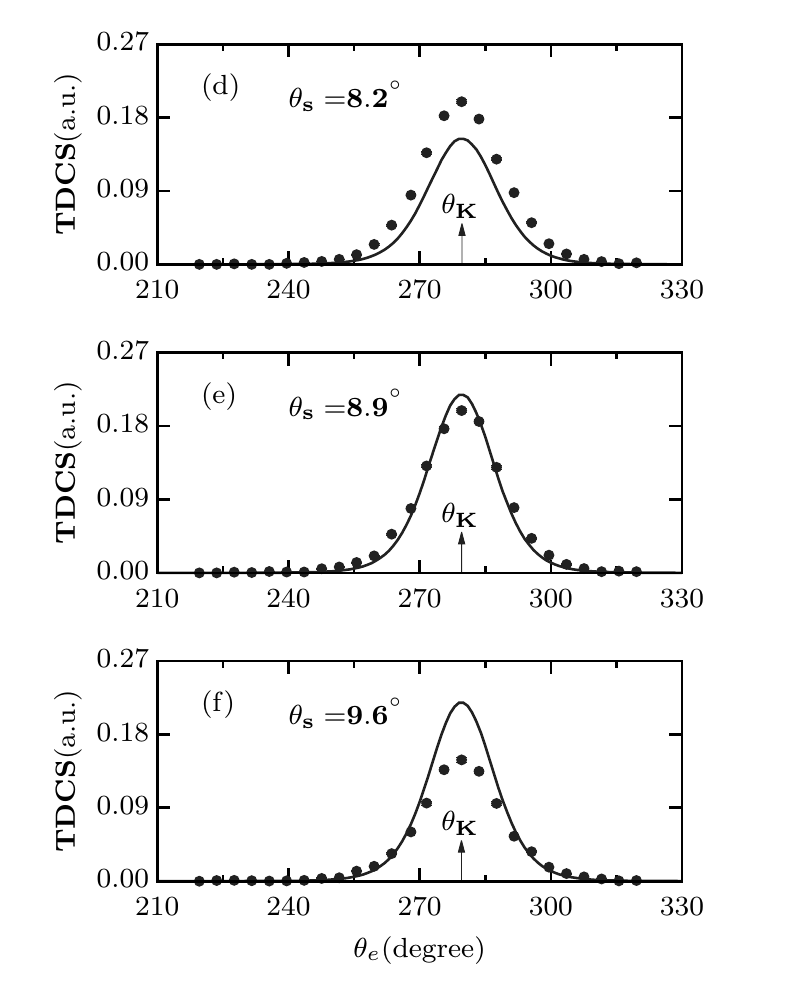}
\end{subfigure}
\caption {TDCS as a function of ejection angle $\theta_e$ for the plane wave (e,2e) process on $H_2$ molecule. Plane wave results are represented by solid line and experimental results (\cite{Cherid1989}) by full circles. Sub-figures (a)-(c) represent the results for an incident energy $E_i$ = 4087 eV, emission energy $E_e$ = 20 eV and scattering angles $\theta_s$ shown in each figure. While, sub-figures (d)-(f) represent the results for an incident energy $E_i$ = 4168 eV, emission energy $E_e$ = 100 eV and scattering angles $\theta_s$ shown in each figure. We have scaled up the theoretical results for sub-figures (d)-(f) by a factor of 10 respectively.  Arrows indicate the direction of momentum transfer $(\theta_\mathbf{K})$ and recoil direction $(\theta_{-\mathbf{K}})$ (opposite to $(\theta_\mathbf{K})$ direction) for this and all subsequent figures.}
\label{fig3}
\end{figure*}

Since it is difficult to obtain the exact alignment of the Bessel beam with the target, one needs to consider the generalized equation for the Bessel beam given by equation (\ref{e19}), such that

\begin{equation}\label{e27}
\psi^{(tw)}_{\varkappa m}(\mathbf{r}_0) = \int^{\infty}_{0} \frac{dk_{i\perp}}{2\pi}\ k_{i\perp} \int_{0}^{2\pi}\frac{d\phi_p}{2\pi}\ a_{\varkappa m}(k_{i\perp})e^{i\mathbf{k}_i \cdot \mathbf{r}_0}e^{-i\mathbf{k}_i \cdot \mathbf{b}},
\end{equation}
where, $\mathbf{b}$ is the vector that describes the transverse orientation of the incident twisted electron beam with respect to the beam direction. The impact parameter vector $\mathbf{b}$ is defined as $\mathbf{b} = b \cos\phi_b \hat{x} + b\sin\phi_b \hat{y}$, where \textit{b} is the magnitude of $\mathbf{b}$ and $\phi_b$ is the azimuthal angle of  $\mathbf{b}$. The additional factor $e^{-i\mathbf{k}_i \cdot \mathbf{b}}$ indicates the complex spatial structure of the Bessel beam contrary to the plane wave \cite{Serbo2015}. We can get equation (\ref{e19}) by just substituting $\mathbf{b}$ = 0 in equation (\ref{e27}). As we did for \textit{b} = 0 case to get the $T_{fi}^{tw}$ (refer to equation (\ref{e24})), we can get the twisted wave transition amplitude $T^{tw}_{fi}$ in terms of plane wave transition amplitude as;
\begin{equation}\label{e28}
T^{tw}_{fi}(\varkappa,\mathbf{K},\mathbf{b}) = (-i)^m \sqrt{\frac{\varkappa}{2\pi}} \int_{0}^{2\pi} \frac{d\phi_p}{2\pi}\ e^{im\phi_p - ik_{i\perp}b}\ T_{fi}(\mathbf{K}).
\end{equation}
where, $k_{i\perp}b$ = $\varkappa b \cos(\phi_p-\phi_b)$.
The scenario wherein a single molecule scatters the incident twisted electron beam is hardly achieved experimentally. It is, therefore, imperative to consider a macroscopic molecular target. The cross-section for such a target can then be computed by taking the average of the plane wave cross-sections over all the possible impact parameters, \textbf{b}, in the transverse plane of the twisted electron beam.

\begin{figure*}
\centering
\begin{subfigure}{.5\textwidth}
  \centering
  \includegraphics[width=1.0\linewidth]{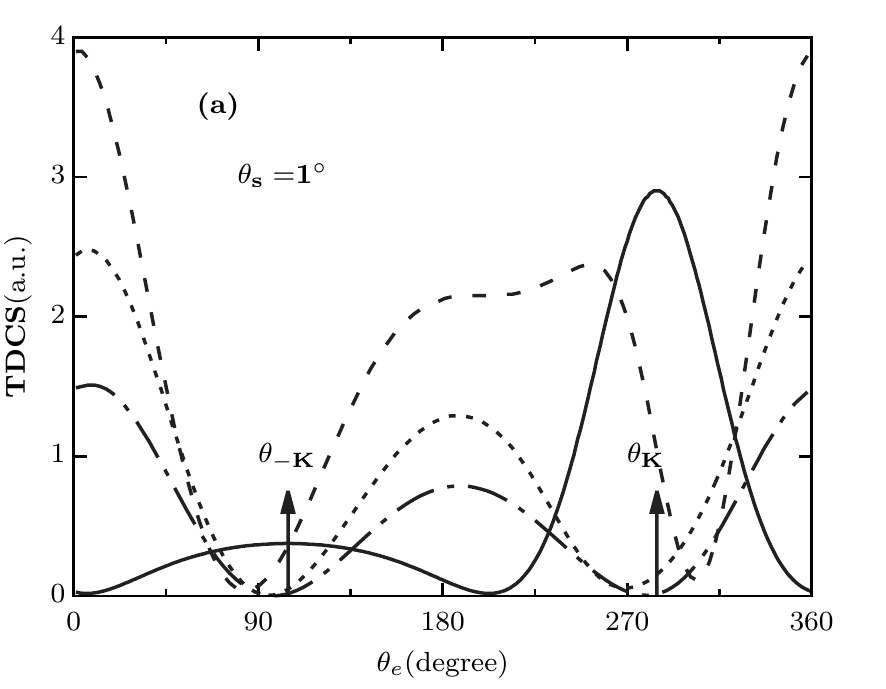}
\end{subfigure}%
\begin{subfigure}{.5\textwidth}
  \centering
  \includegraphics[width=1.0\linewidth]{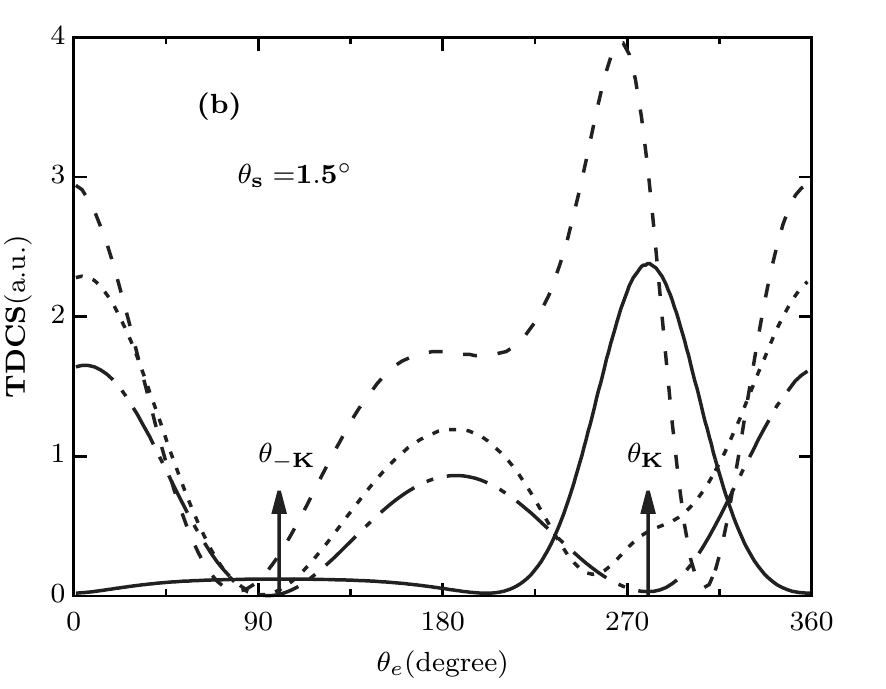}
\end{subfigure}
\caption{TDCS as a function of ejection angle $\theta_e$ for the twisted electron (e,2e) process on $H_2$ molecule in the co-planar asymmetric geometry. The kinematics used here is $E_i$ = 4087 eV, $E_e$ = 20eV and scattering angles (a) $\theta_s$ = 1\textdegree \ and (b) 1.5\textdegree. We keep $\theta_p$ = $\theta_s$ here. The solid, dashed, dotted and dashed-dotted curves represent the plane wave, \textit{m} = 1,2 and 3 respectively. The results for $m \neq 0$ are scaled up by a factor of 10 in both the sub-figures (a) and (b).}\label{fig4}
\end{figure*}

 The average cross-section, $(TDCS)_{av} = \overline{\frac{d^3 \sigma}{d\Omega_s d\Omega_e dE_e}}$ in terms of plane wave cross-section can be described as (for detailed derivation see Harris {\it et al.}(2019) \cite{Harris2019});
\begin{equation}\label{e29}
(TDCS)_{av} = \frac{1}{2 \pi \cos\theta_p} \int_{0}^{2\pi} d\phi_p \frac{d^3 \sigma(\mathbf{K})}{d\Omega_s d\Omega_e dE_e},
\end{equation}
where $\frac{d^3 \sigma(\mathbf{K})}{d\Omega_s d\Omega_e dE_e}$ is like the TDCS for the plane wave electron beam depending on \textbf{K}.
From equation (\ref{e29}), it is evident that the TDCS averaged over the impact parameter is independent of the OAM number {\it m} of the incident twisted electron beam.

\section{Results and discussion}\label{sec3}

\subsection{Angular profiles of the TDCS for plane wave}

\begin{figure*}
\centering
\begin{subfigure}{.5\textwidth}
  \centering
  \includegraphics[width=1.0\linewidth]{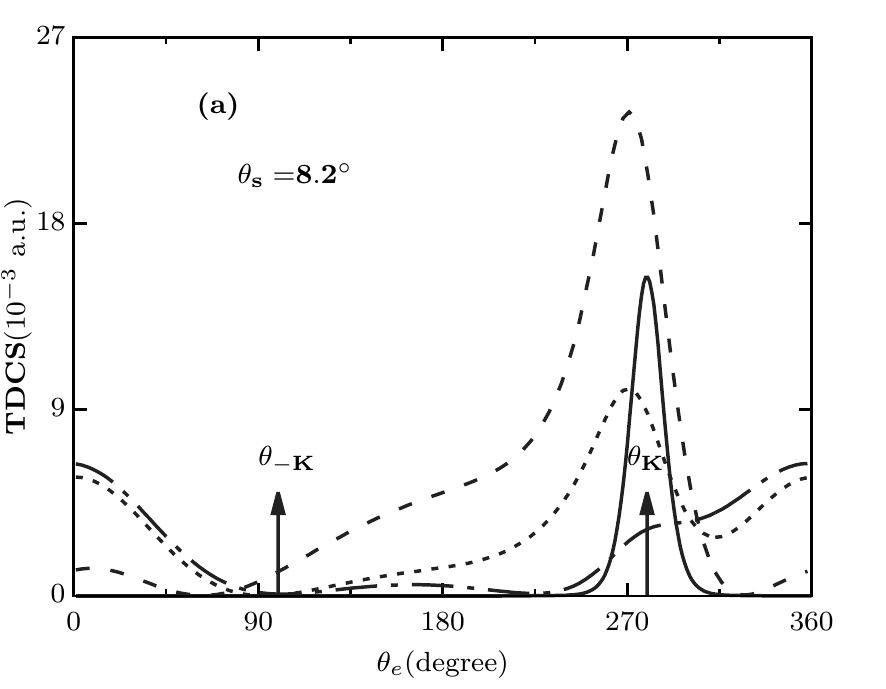}
\end{subfigure}%
\begin{subfigure}{.5\textwidth}
  \centering
  \includegraphics[width=1.0\linewidth]{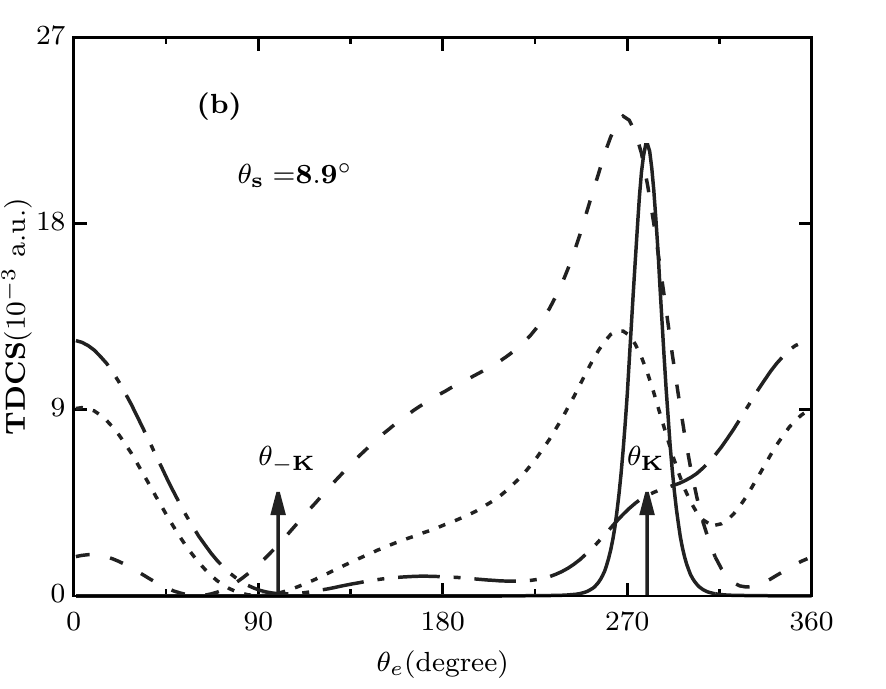}
\end{subfigure}
\caption{Same as figure \ref{fig4} except for the kinematics $E_i$ = 4168 eV, $E_e$ = 100 eV and scattering angles (a) $\theta_s$ = 8.2\textdegree \ and (b) 8.9\textdegree. We keep $\theta_p$ = $\theta_s$ here. The results for $m \neq 0$ are scaled up by a factor of 1000 in the sub-figure (a) and by 2000 in sub-figure (b).}\label{fig5}
\end{figure*}

In the figures \ref{fig3}-\ref{fig5}, we present the results of our calculations of the TDCS as a function of the ejected electron angle in the coplanar asymmetric geometry for different scattering angles (given in the frame of each figure). We use the same kinematics as used by Ch\'erid \textit{et al.} \cite{Cherid1989} wherein they took the incident energy as 4087 eV and 4168 eV. We are using the first Born approximation because the computation of the TDCS with other approaches is tedious and gets more complicated when we compute TDCS for the twisted electron (e,2e) process. Also, our first Born approximation model is analytic in the computation of $T_{fi}$, which makes our computation of $T_{fi}^{tw}$ easier as it requires only the integration over the azimuthal angle, $\phi_p$, of the twisted electron. One can perform the integration numerically. To the best of our knowledge, there has been no detailed study on the (e,2e) processes on the $H_2$ molecule for the twisted electron incidence case. Hence, our results in this paper should be taken from that perspective. We are sure that the present study will advance the theoretical and experimental research in this field. In literature, more sophisticated models exist (like, two-effective center (TEC), molecular three-body distorted wave (M3DW), multicenter molecular three-body distorted-wave (MCM3DW), multicenter distorted-wave (MCDW) etc.) for the plane wave (e,2e) processes on $H_2$ molecule \cite{Weck2001,Gao2006,Li2018,Ali2019}. We hope that one can also explore the twisted (e,2e) process with a more sophisticated model for further grinding of our results in the future.

\begin{figure*}
\centering
\begin{subfigure}{.5\textwidth}
  \centering
  \includegraphics[width=1.1\linewidth]{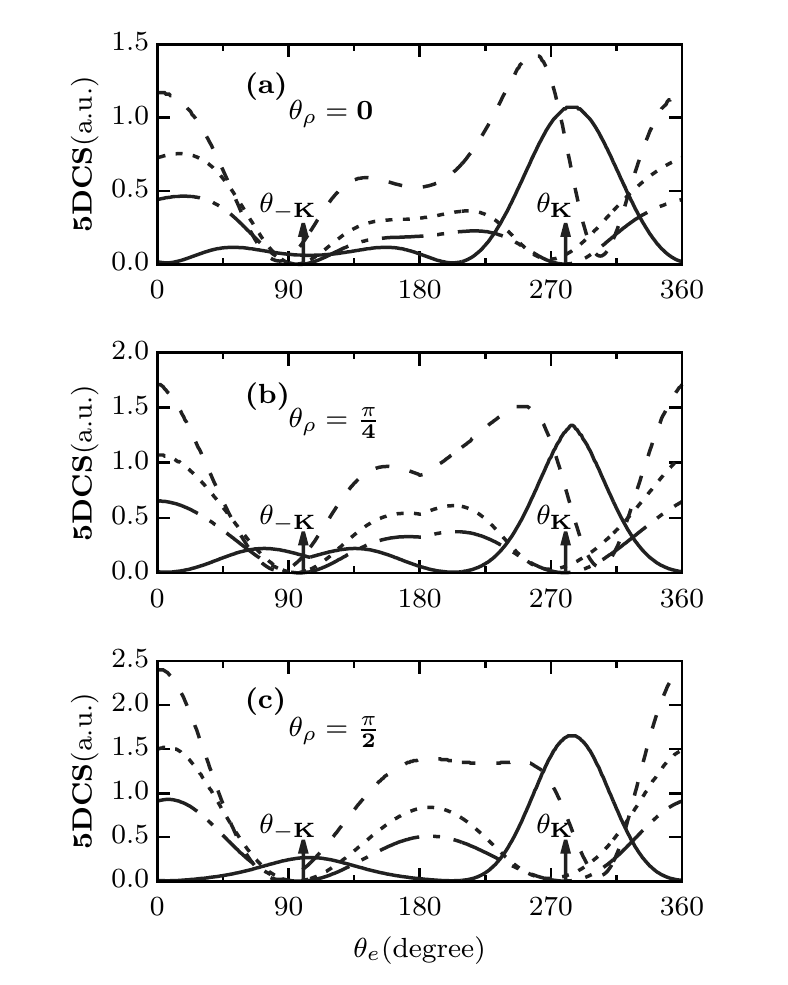}
\end{subfigure}%
\begin{subfigure}{.5\textwidth}
  \centering
  \includegraphics[width=1.1\linewidth]{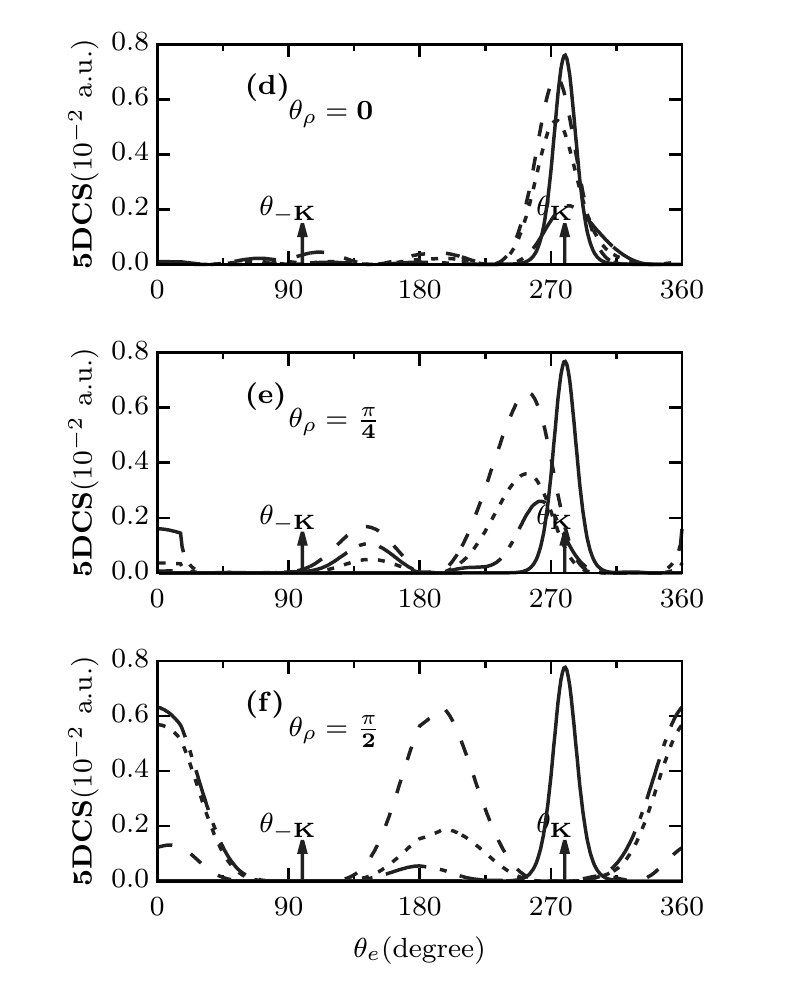}
\end{subfigure}
\caption{5DCS as a function of ejection angle $\theta_e$ for the twisted electron incidence for different orientations $\boldsymbol{\rho}$ of the molecule in the coplanar asymmetric geometry.  Sub-figures (a)-(c) represent the results for an incident energy $E_i$ = 4087 eV, emission energy $E_e$ = 20 eV, scattering angle $\theta_s$ = 1\textdegree \ and orientation angle, $\theta_{\rho}$ shown in each figure. While, sub-figures (d)-(f) represent the results for $E_i$ = 4168 eV, $E_e$ = 100 eV, $\theta_s$ = 8.2\textdegree \ and $\theta_{\rho}$ shown in each figure. In both the kinematics, we keep $\theta_p$ = $\theta_s$. The magnitude of the 5DCS is scaled up by a factor of 10  for figure \ref{fig6}(a)-(c).  For figure \ref{fig6}(d)-(e) we use different scaling factors, {\it viz.} 250, 450 and 550 for {\it m} = 1,2 and 3 respectively for $\theta_{\rho}$ = 0;  400, 500 and 2000 for {\it m} = 1, 2 and 3 respectively for $\theta_{\rho}$ =  $\pi/4$ and 900 for all three values of {\it m} for $\theta_{\rho}$ = $\pi/2$.}
\label{fig6}
\end{figure*}
 We present in figure \ref{fig3}, the results of our calculations of TDCS as a function of the angle of the ejected electron ($\theta_e$) for the plane wave electron beam in the coplanar asymmetric geometry. The arrows used in the figure \ref{fig3} and subsequent figures represent the direction of momentum transfer ($\theta_\mathbf{K}$) and the recoil direction ($\theta_{-\mathbf{K}}$), opposite to $\theta_{\mathbf{K}}$. We take TDCS, as experimentally it is difficult to align the molecule in a particular direction. As explained in equation (\ref{e16}), TDCS is computed by integrating 5DCS over the angular alignments of the $H_2$ molecule.  We keep $E_i$ = 4087 eV, $E_e$ = 20 eV and different $\theta_s$ as shown in the figure \ref{fig3}(a)-(c) and $E_i$ = 4168 eV, $E_e$ = 100 eV and different $\theta_s$ as shown in the figure \ref{fig3}(d)-(f). 

	We compare the results of our calculations with the experimental data \cite{Cherid1989}. At this kinematics, we can justify our theoretical model based on the following arguments;
\begin{enumerate}
\item Because of the high energy of the incident and the scattered electrons, the selection of the plane wave function for both the electrons is appropriate.
\item  The ejected electron carries lower energy. The choice of the Coulomb wave function for the ejected electron is also justified.
\end{enumerate}

\begin{figure*}
\centering
\begin{subfigure}{.5\textwidth}
  \centering
  \includegraphics[width=1.1\linewidth]{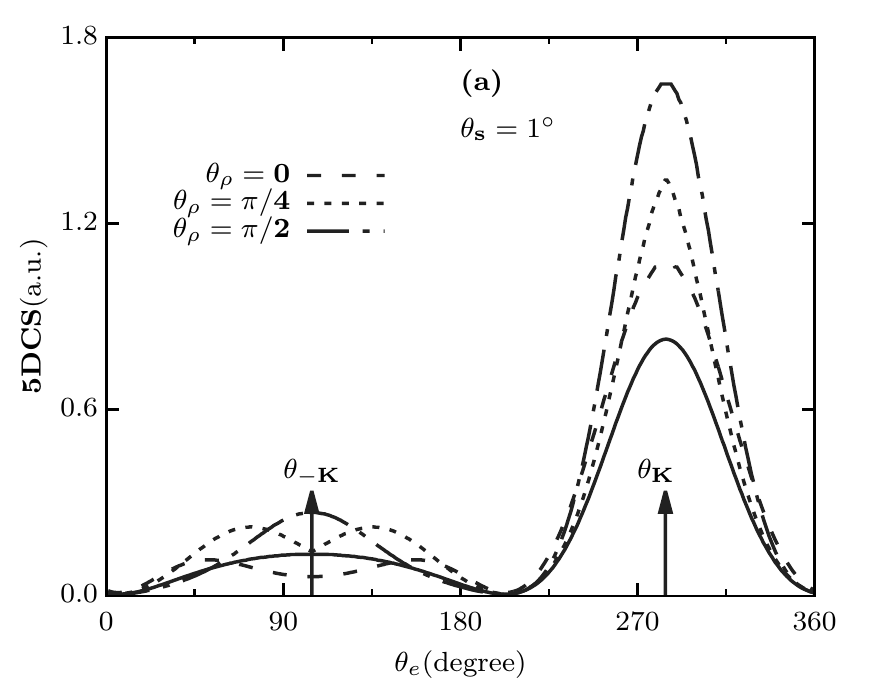}
\end{subfigure}%
\begin{subfigure}{.5\textwidth}
  \centering
  \includegraphics[width=1.1\linewidth]{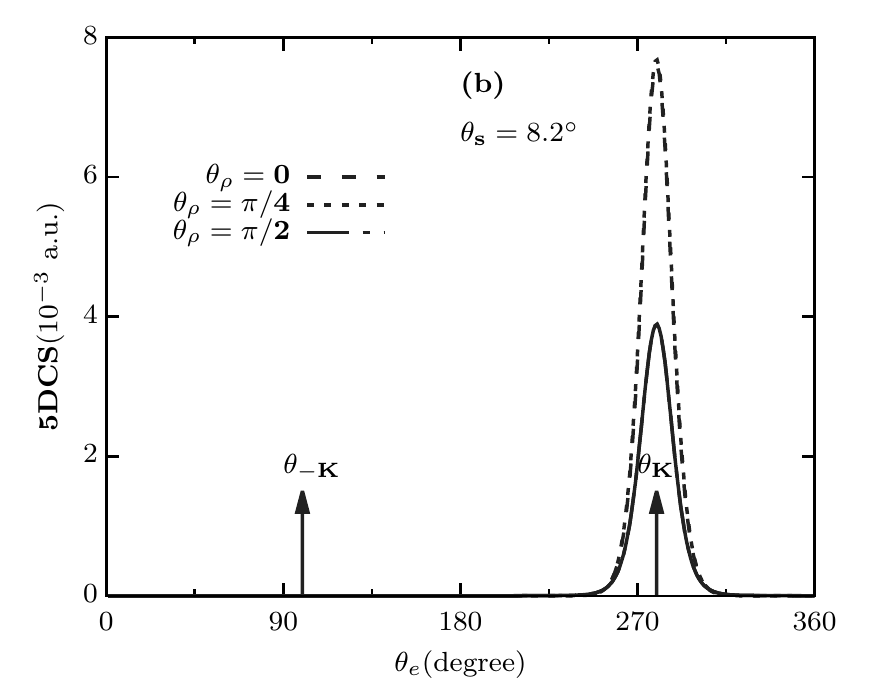}
\end{subfigure}
\caption{5DCS with and without interference as a function of ejection angle $\theta_e$ for the twisted electron incidence for different orientations $\boldsymbol{\rho}$ of the molecule in the coplanar asymmetric geometry. The solid curve represents the 5DCS without interference term. The dashed, dotted and dashed-dotted curves represent the 5DCS with interference term for different orientations as shown in each sub-figure. We used the kinematics $E_i$ = 4087 eV, $E_e$ = 20 eV and $\theta_s$ = 1\textdegree\ for sub-figure (a). Sub-figure (b) represents the 5DCS for the kinematics $E_i$ = 4168 eV, $E_e$ = 100 eV and $\theta_s$ = 8.2\textdegree.}
\label{fig7}
\end{figure*}

 We compare our results of TDCS with the experimental data to benchmark our calculation so that we can justify our calculations for the twisted electron (e,2e) process.  From the comparison, we found that the present theoretical model reproduces the experimental trends for most cases. Our calculations reproduce the prominent binary peak and the shallow recoil peak, wherever appears (see figure \ref{fig3}).
We want to add here that to compare our calculations with the experimental data, we have scaled up our theoretical calculations for figures \ref{fig3}(d)-(f) by a factor of 10. For figures \ref{fig3}(a)-(c), no normalization factor is used in our calculation to compare them with the experimental data. 

\begin{figure*}
  \includegraphics[width=2.2\columnwidth]{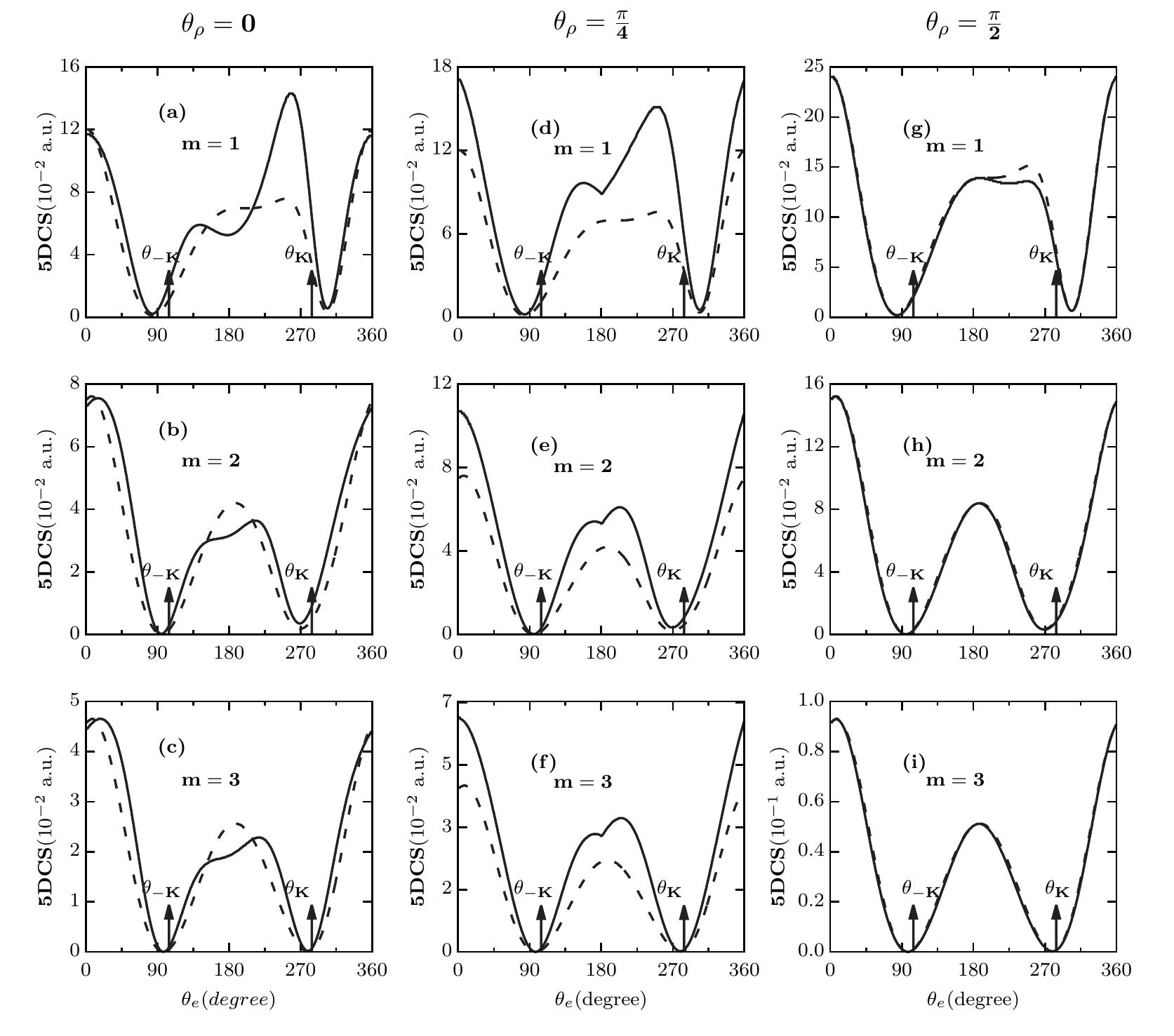}
\caption{Same as figure \ref{fig7} except that the 5DCSs are plotted for the twisted electron beam for different {\it m} and $\theta_{\rho}$ as shown in the figure for the kinematics $E_i$ = 4087 eV, $E_e$ = 20 eV and $\theta_s$ = $\theta_p$ = 1\textdegree in the coplanar asymmetric geometry.}
\label{fig8}
\end{figure*}

\subsection{Angular profiles of the TDCS for twisted electron beam}

Having benchmarked our calculations for the plane wave with the experimental data, we present the results of our calculations of TDCS with the twisted electron beam. It is worth adding that we have also benchmarked the elastic scattering cross-section on a $H_2$ molecule with the reported results by \cite{Maiorova2018}. We were able to reproduce the results predicted by Maiorova {\it et al.} (2018) \cite{Maiorova2018}. We are not presenting the results of our calculations of the elastic cross-section for $H_2$ molecule for brevity. We have also compared our calculation of the (e,2e) process on the hydrogen atom for the twisted electron with the one reported by Harris \textit{et al.} (2019) \cite{Mandal2020} treating the ejected electron as plane wave. We confirmed that our calculations also reproduce the results of Harris \textit{et al.} for an impact energy, $E_i$ = 1 keV and $E_e$ as 20 eV, 50 eV and 100 eV for $\theta_s$ = 10 mrad with different values of $\theta_p$ , i.e. 1 mrad, 10 mrad, 100 mrad and {\it m} = 1, except the contrasting results in the magnitude of TDCS which is not a major concern here. This maybe due to the choice of the Coulomb wave function for the ejected electron in our model.

 We present the results of our calculations of TDCS for the twisted electron incidence for $E_i$ = 4087 eV in figure \ref{fig4} for $\theta_s$ = 1.0\textdegree \ and 1.5\textdegree. We keep $\theta_p$ = $\theta_s$ and vary {\it m} from 1 to 3 in steps of 1. In figure \ref{fig4} and all the subsequent figures we represent the solid, dashed, dotted and dashed-dotted curves  for plane wave, {\it m} = 1,2 and 3 respectively, unless otherwise stated. We found that the magnitude of TDCS for {\it m} $\neq$ 0 is reduced by at least an order of 1 when compared with {\it m} = 0 (the plane wave calculations). For example, the peak value of TDCS for the plane wave is 2.9 a.u whereas for {\it m} = 1 case, it is around 2.37 a.u.(multiplied by a factor of 10 to compare it with the plane wave calculation (see solid and dashed curves in figure \ref{fig4}(a))). This is further reduced when we gradually increase {\it m} upto 3. We also observe that for the twisted electron beam calculation, the binary peak largely disappears for \textit{m} = 2 and \textit{m} = 3 (see dashed and dashed-dotted curves in the region marked by arrow in the binary region in figure \ref{fig4}). We also found substantial contribution in TDCS for forward and backward direction for {\it m} = 1,2 and 3 (see peak around $\theta_e = $0\textdegree \ and 180\textdegree \ for dashed, dotted and dashed-dotted curves). We further observe that only for \textit{m} = 1 case, there is a binary peak, which is shifted to smaller angle (see dashed curves in figure \ref{fig4}). This clearly signifies the effect of different values of {\it m} on the angular profile of the TDCS.
 
	In the previous kinematics, we have seen that how the angular profiles of TDCS depend on \textit{m} for small momentum transfer. We now discuss the angular profile of TDCS for the other kinematics ($E_i$ = 4168eV, $E_1$ = 100eV, $\theta_s$ = 8.2\textdegree\ and 8.9\textdegree  ) and $\theta_p$ = $\theta_s$ for different values of \textit{m} in figure \ref{fig5} to further investigate the effect of the larger momentum transfer on the angular profiles of the TDCS. We plot the angular profile of TDCS for $\theta_p$ = $\theta_s$ = 8.2\textdegree\ in figure \ref{fig5}(a) and for $\theta_p$ = $\theta_s$ = 8.9\textdegree\ in figure \ref{fig5}(b). In this kinematics, since there is a large momentum transfer as compared to the previous kinematics, we find the smaller cross sections of TDCS. The order of cross section is $10^{-5}$ which is significantly smaller than the previous case (see figure \ref{fig4}). For all the values of \textit{m} for $\theta_s$ = 8.2\textdegree\ (figure \ref{fig5}(a)), we observe the binary peak as the dominant peak with peak position shifted from the  momentum transfer direction $\theta_{\mathbf{K}}$ (see the arrow marked at $\theta_e$ = $\theta_{\mathbf{K}}$ in the figure \ref{fig5}). Like as observed in the previous kinematics (refer to figure \ref{fig4}), we also observe peak in the forward direction for the different values of \textit{m} (see dashed, dotted and dashed-dotted curves around $\theta$ = 0\textdegree\ (or 360\textdegree) region in the figure \ref{fig5}(a)). However, in this case the intensity of the peak is not as pronounced as that for the previous case (see dashed, dotted and dashed-dotted curves in figure \ref{fig4} and \ref{fig5} for comparison). In fact, for \textit{m} = 1, there is hardly any substantial forward peak when compared with its binary peak (see dashed curve in figure \ref{fig5}). As \textit{m} increases, the forward peak gets enhanced. In fact, the forward peak  for {\it m} = 3 dominates over the shoulder structure in the binary peak region (see dashed-dotted curve in figure \ref{fig5}). When we observe the angular profile of the TDCS for $\theta_s$ = 8.9\textdegree, we find that the magnitude of the TDCS is further reduced which is, of course, expected as in this case we have a bit larger momentum transfer as compared to $\theta_s$ = 8.2\textdegree\ case. In addition to these observations, we find that the backward peak is slightly enhanced for $m \neq 0$, especially for \textit{m} = 1 (see dashed curve around $\theta_e$ = 180\textdegree\ in figure \ref{fig5}(b) and compare it with that in the figure \ref{fig5}(a)).
\begin{figure*}
  \includegraphics[width=2.2\columnwidth]{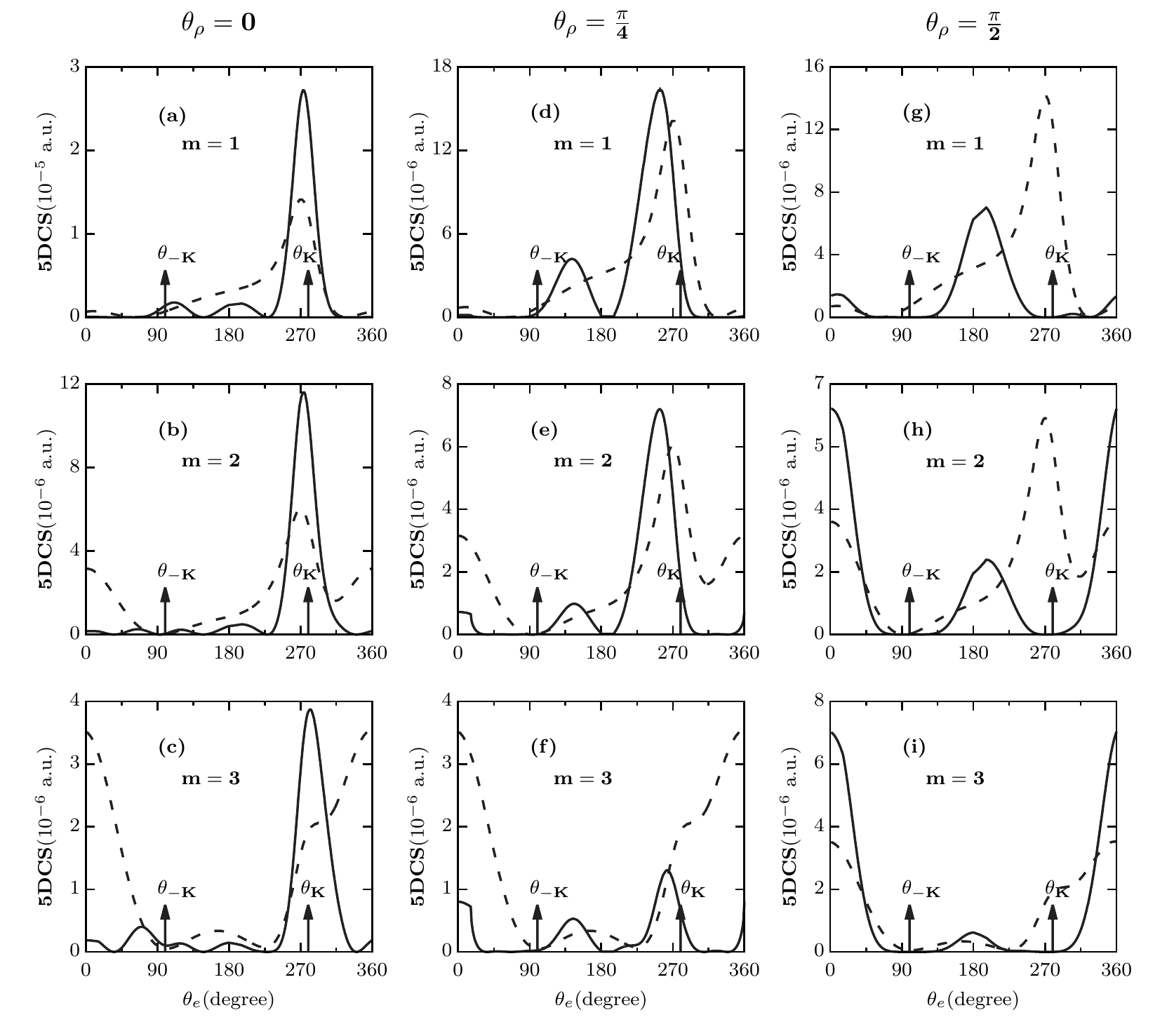}
\caption{Same as figure \ref{fig8} except for the kinematics $E_i$ = 4168 eV, $E_e$ = 100 eV and $\theta_s$ = $\theta_p$ = 8.2\textdegree\ in the coplanar asymmetric geometry.}
\label{fig9}
\end{figure*}

\subsection{Orientation effects on the TDCS}

 Till now, we studied the angular profiles of the TDCS averaged over all the possible orientations of the inter-nuclear distance $\boldsymbol{\rho}$ of $H_2$ molecule since experimentally it is difficult to orient the molecule in some preferred direction. However, we can still theoretically investigate the orientation effects on the 5DCS by exploring the angular distribution of 5DCS for some fixed orientation of the $H_2$ molecule to get more details about (e,2e) process. We computed the  TDCS for different $\theta_{\rho}$, which is the angle between the incident electron and the molecular axis of the $H_2$ molecule, in the coplanar geometry ($\phi_{\rho}$ = 0). 
 
 Figure \ref{fig6} represents the angular profile of the 5DCS with respect to the $\theta_e$ for three different orientations, {\it viz.} ($\theta_{\rho}, \phi_{\rho}$) = (0, 0), ($\frac{\pi}{4}$,0) and ($\frac{\pi}{2}$,0) of the $H_2$ molecule.  
 We choose the same kinematics as used in the figure \ref{fig3}(a) with $\theta_s$ = $\theta_p$ = 1.0\textdegree\ for figure \ref{fig6}(a)-(c). While, for figure \ref{fig6}(d)-(f), we choose the kinematics used in figure \ref{fig3}(d) for $\theta_s$ = $\theta_p$ = 8.2\textdegree. The orientation (0,0) corresponds to the alignment of the $H_2$ molecule along the beam direction whereas the ($\frac{\pi}{2}$,0) corresponds to the transverse orientation of the $H_2$ to the beam direction. The orientation ($\frac{\pi}{4}$,0) corresponds to the intermediate of the above mentioned orientations. We scale up the magnitude of the 5DCS for {\it m} $\neq$ 0 case by a factor of 10 for the figure \ref{fig6}(a)-(c) and for figure \ref{fig6}(d)-(f), we use different scaling factors to compare them with the plane wave results.  For the parallel orientation, $\theta_{\rho}$ = 0, we use the factors 250, 450 and 550 for {\it m} = 1, 2 and 3 respectively. The results for perpendicular orientation, $\theta_{\rho} = \pi/2$, are scaled up by factors 400, 500 and 2000 for {\it m} = 1, 2 and 3 respectively and for $\theta_{\rho} = \pi/4$  we use factor 900 for all {\it m}'s. 
\begin{figure*}
\centering
\begin{subfigure}{.5\textwidth}
  \centering
  \includegraphics[width=1.0\linewidth]{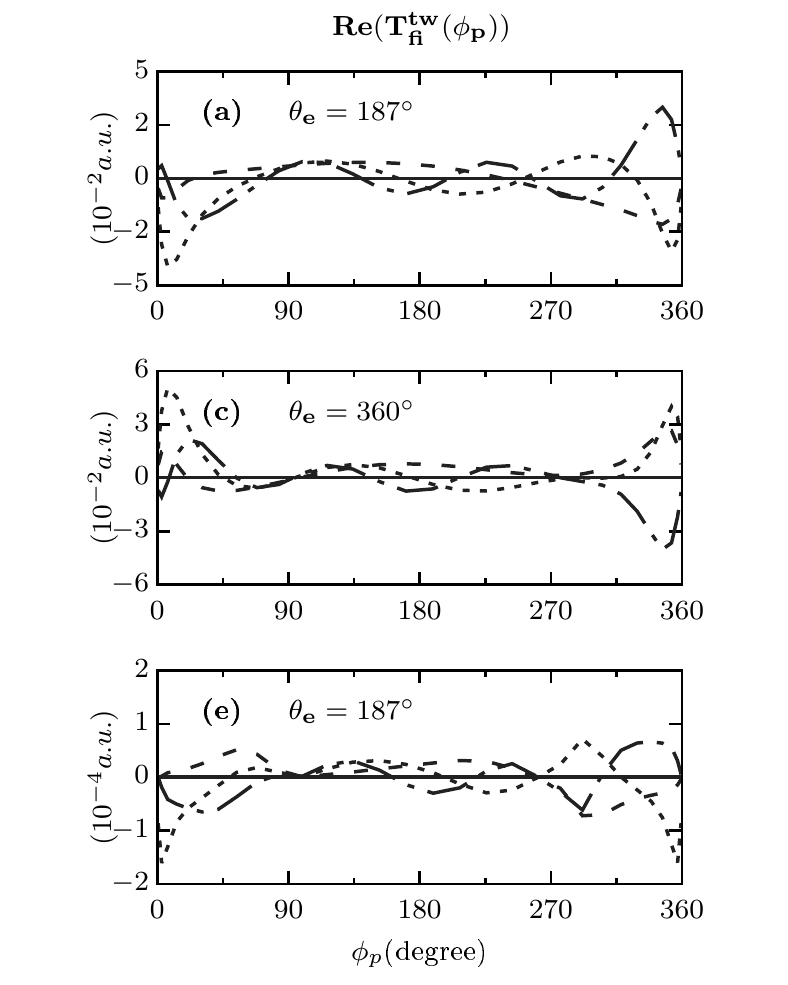}
\end{subfigure}%
\begin{subfigure}{.5\textwidth}
  \centering
  \includegraphics[width=1.0\linewidth]{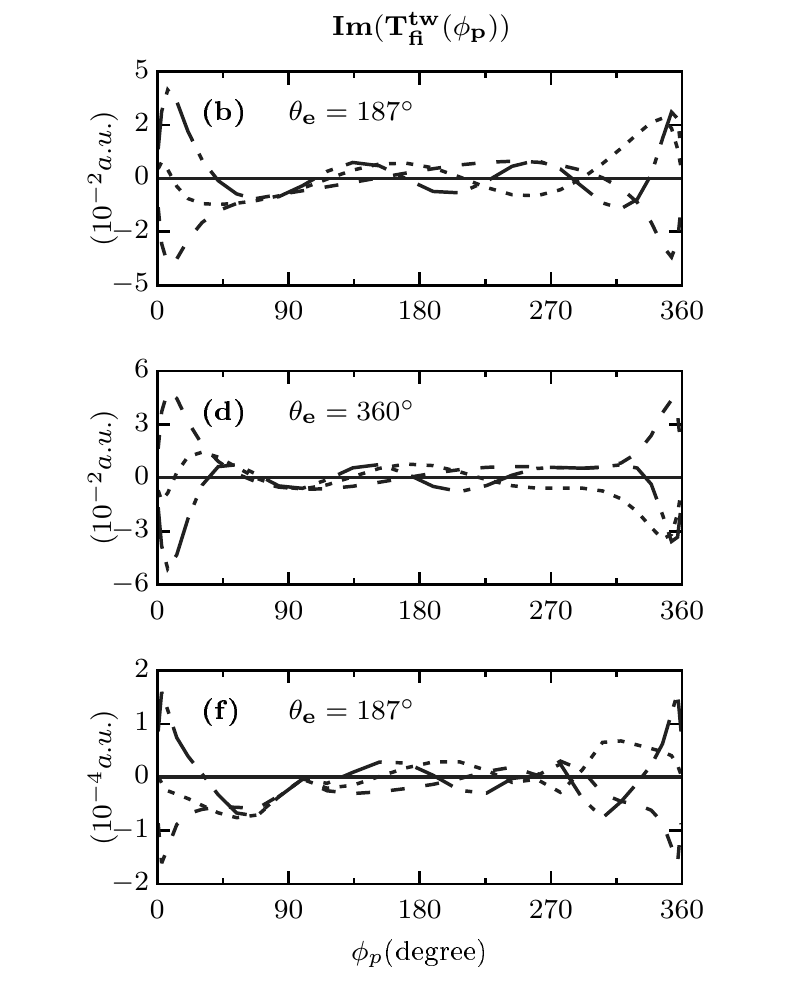}
\end{subfigure}
\caption{Real and imaginary parts of the transition amplitude ($T_{fi}^{tw}(\phi_p)$) as a function of the azimuthal angle ($\phi_p$) of the twisted electron beam. The kinematics for sub-figures (a)-(d) is $E_i$ = 4087eV, $E_1$ = 200eV, $\theta_s$ = $\theta_p$ = 1\textdegree, $\theta_{\rho}$ = $\frac{\pi}{2}$ and $\theta_e$ given in each frame. For the sub-figures (e)-(f) the kinematics used is $E_i$ = 4168eV, $E_1$ = 100eV, $\theta_s$ = $\theta_p$ = 8.2\textdegree, $\theta_{\rho}$ = $\frac{\pi}{2}$ and $\theta_e$ given in frame. The dashed, dotted and dashed-dotted curves represent the Re($T_{fi}^{tw}(\phi_p)$) and Im($T_{fi}^{tw}(\phi_p)$) for {\it m} = 1, 2 and 3 respectively.}
\label{fig10}
\end{figure*}

In figure \ref{fig6}(a)-(c), for all the orientations of $\boldsymbol{\rho}$, we observe peak in the forward direction (see dashed, dotted and dashed-dotted curves in figure \ref{fig6}(a)-(c) in the region $\theta_e$ = 0\textdegree (or 360\textdegree). In addition to this, we  also observe a prominent peak in the backward direction for $\theta_{\rho} = \frac{\pi}{2} $(see dashed, dotted and dashed-dotted curve in the figure \ref{fig6}(c) in the $\theta_e$ = 180\textdegree\ region) for different values of {\it m}. For other orientations, we do not observe peak in the backward region (see figure \ref{fig6}(a)-(b) around $\theta_e$ = 180\textdegree region). In fact, the 5DCS contributions in the backward region increases with $\theta_{\rho}$ as can be observed from the figures \ref{fig6}(a)-(c) in the backward region.   Further, we observe that only for {\it m} = 1, we observe a shifted binary peak in the angular profile of 5DCS (see dashed curve around the region marked by arrow for $\theta_e$ = 273\textdegree) for all the orientations of $\boldsymbol{\rho}$.
  
 For figures \ref{fig6}(d)-(f), we observe a prominent peak for the parallel orientation ($\theta_{\rho} = 0$) in the binary region for all the values of {\it m} (see dashed, dotted and dashed-dotted curve in figure \ref{fig6}(d)). Furthermore, we observe that the binary peak for {\it m} = 1 and 2 shifts from the momentum transfer direction (see dashed and dotted curves in the figure \ref{fig6}(d) around $\theta_e$ = $\theta_{\mathbf{K}}$ region). In the recoil region, we observe a very small contribution to the differential cross-section from all the values of {\it m}. For $\theta_{\rho}$ = $\frac{\pi}{4}$, we observe peaks in binary and recoil regions for {\it m} = 1, 2 and 3, which are shifted significantly from the momentum transfer direction (see dashed, dotted and dashed-dotted curve in figure \ref{fig6}(e) around the region marked by the arrows). For the perpendicular orientation ($\theta_{\rho} = \frac{\pi}{2}$), we observe prominent peaks in both the forward and backward regions (see dashed, dotted and dashed-dotted curve in figure \ref{fig6}(f) around $\theta_e$ = 0\textdegree (or 360\textdegree) and 180\textdegree). For all the orientations of $\boldsymbol{\rho}$, the magnitude of the 5DCS drops significantly for higher {\it m}, in both the kinematical arrangements. From the above discussion, it is evident that there is a substantial effect of the OAM number {\it m}, on the angular profiles of 5DCS for different $\theta_{\rho}$.

\subsection{Interference effects}

The study of the ionization of $H_2$ molecule provides an opportunity to investigate the Young-type interference-effect in the (e,2e) process. The interference effect is attributed to the scattering process of the incident electron from the two identical H-atoms of the $H_2$ molecule. In figure \ref{fig7}, we investigate the interference effects in the (e,2e) process for different orientations of the  $H_2$ molecule for plane wave electron beam. We choose the same $\theta_{\rho}$, as used in figure \ref{fig6}. We use solid, dashed, dotted and dashed-dotted curves respectively for 5DCS without interference, 5DCS with interference for $\theta_{\rho}$ = 0, $\frac{\pi}{4}$ and $\frac{\pi}{2}$. We use the same kinematics as used for figure \ref{fig4}(a) and figure \ref{fig5}(a) for figure \ref{fig7}(a) and (b) respectively. We observe binary and recoil peaks in the 5DCS without interference for all the orientations (see the solid curve in the figure \ref{fig7}(a)). However, when we take into account the interference term in the 5DCS, we observe that the binary peak is enhanced for all the orientations, which is due to the constructive interference of the two H-atom scattering (see dashed, dotted and dashed-dotted curves in the figure \ref{fig7}(a)). In the recoil region, for $\theta_{\rho}$ = 0 orientation, there is a destructive interference in the recoil peak region leading to a dip in the 5DCS around $\theta_{-\mathbf{K}}$ region (see dashed curve and compare it with the solid curve in the $\theta_e$ = $\theta_{-\mathbf{K}}$ region). As we increase $\theta_{\rho}$ from 0 to $\frac{\pi}{2}$, the interference pattern converts from the destructive to constructive type which is reflected in the enhancement of the recoil peak for $\theta_{\rho}$ = $\frac{\pi}{2}$ (see dashed-dotted curve in figure \ref{fig7}(a)). Thus, it is obvious that the cross-section is maximum for the perpendicular orientation (see the dashed-dotted curve in figure \ref{fig7}).
Further, when we use a multiplicative factor 2 for the 5DCS without interference, we find that the binary and recoil peaks coincide for $\theta_{\rho} = \frac{\pi}{2}$ (see solid and dashed-dotted curves in figure \ref{fig7}(a)). In the figure \ref{fig8}(a), the results for $\theta_s$ = 8.2\textdegree\ are quite intriguing, as for this kinematics, we observe the identical angular profile for all the three orientations. Also, when multiplied by a factor of 2, the peak value of the 5DCS without interference is identical to the 5DCS with   term. We observe a binary peak at the same position for all the orientations due to the large momentum transfer for this kinematics. Due to the small role played by the nucleus in the scattering process for this kinematics, the 5DCS is found to be independent of the molecular orientation (see figure \ref{fig7}(b)).

\begin{figure*}
\centering
\begin{subfigure}{.5\textwidth}
  \centering
  \includegraphics[width=1.0\linewidth]{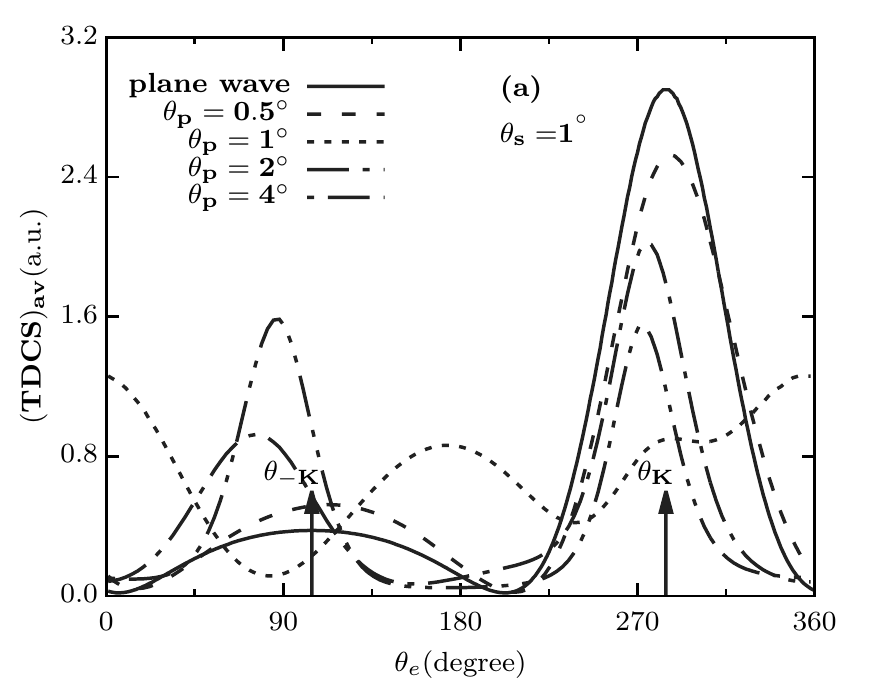}
\end{subfigure}%
\begin{subfigure}{.5\textwidth}
  \centering
  \includegraphics[width=1.0\linewidth]{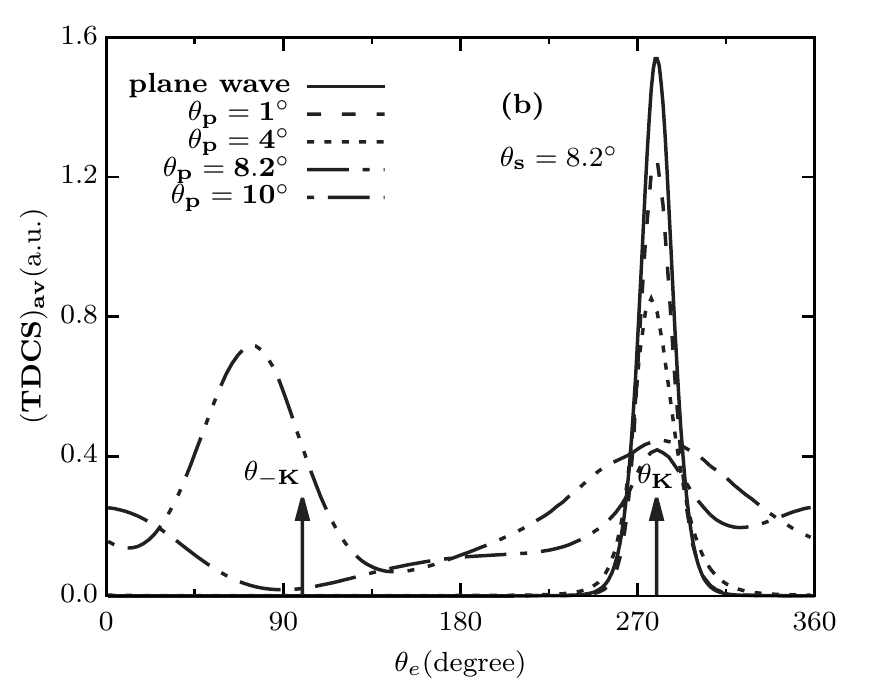}
\end{subfigure}
\caption{TDCS integrated over the impact parameter $\mathbf{b}$ for the ionization of $H_2$ molecule for plane wave electron beam(solid curve) and twisted electron electron beam for different opening angles as shown in the frames of each sub-figure. We used the kinematics $E_i$ = 4087 eV, $E_e$ = 20 eV and $\theta_s$ = 1\textdegree\ for figure \ref{fig11}(a). Figure \ref{fig11}(b) represents the $(TDCS)_{av}$ for the kinematics $E_i$ = 4168 eV, $E_e$ = 100 eV and $\theta_s$ = 8.2\textdegree.}
\label{fig11}
\end{figure*}

	Having studied the interference effects on the 5DCS for plane wave electron beam, for different orientations of the inter-nuclear distance $\boldsymbol{\rho}$, we present the results for the interference effects on the 5DCS for twisted electron for different $\theta_{\rho}$ in figure \ref{fig8} and \ref{fig9}. The solid and dashed curve represent the 5DCS with interference and without interference, respectively, in both the figures \ref{fig8} and \ref{fig9}. In figure \ref{fig8}, we present the results of the 5DCS with interference and without interference for different orientations and different OAM numbers {\it m}, as shown in the figure, for the same kinematics as used for figure \ref{fig4}(a). Due to the constructive interference for {\it m} = 1 case for  $\theta_{\rho}$ = 0, the magnitude of the 5DCS  with interference is larger than that for the 5DCS without interference in the binary peak region (see solid and dashed curves in figure \ref{fig8}(a)). For the three values of {\it m}, we observe that the peaks in the forward region are at the same positions for both the 5DCS types (see figure \ref{fig8}(a)-(c) in the forward region). Due to the destructive interference for $m \neq$ 0, the magnitude of 5DCS with interference in the backward region is smaller than that of the without interference (see solid and dashed curves in figure \ref{fig8}(a)-(c)). For the intermediate orientation, $\theta_{\rho} = \frac{\pi}{4}$, we observe that the interference is constructive as the magnitude of the 5DCS with interference is larger than that for the without interference (see solid and dashed curves for {\it m} = 1, 2 and 3 for $\theta_{\rho} = \frac{\pi}{4}$ in figure \ref{fig8}(d)-(f)). For the perpendicular orientation, $\theta_{\rho} = \frac{\pi}{2}$, we observe that when we scale up the 5DCS without interference by a factor of 2, the angular profile is precisely like the angular profile of the TDCS with interference (see the solid and dashed curve in figure \ref{fig8}(g)-(i) for $\theta_{\rho} = \frac{\pi}{2}$).

	In figure \ref{fig9}, we study the interference effects on the 5DCS for the kinematics, as used in figure \ref{fig5}(a) for different orientations of $\boldsymbol{\rho}$. For $\theta_{\rho}$ = 0, the oscillations in the angular profiles of the 5DCS with interference term are more pronounced for larger value of the OAM number, {\it m} (see the solid curve in the figure \ref{fig9}(a)-(c)). For $\theta_{\rho} = \frac{\pi}{4}$, we observe a two to three peak structure for the 5DCS with interference for the three values of {\it m}. However, for the angular profiles of the 5DCS without interference we observe a two peak structure (see solid and dashed curves  in figure \ref{fig9}(d)-(f)). For the perpendicular orientation ($\theta_{\rho} = \frac{\pi}{2}$), we observe peaks in the forward and backward regions for all the {\it m}'s in the angular profile of the 5DCS with interference (see solid curves in figure \ref{fig9}(g)-(i)). In addition to this, we observe prominent peaks in the binary region for {\it m} = 1 and 2 for the angular profile of the TDCS without interference for the same orientation. On the inclusion of the interference term, we observe a destructive interference in the binary peak region (see dip in the solid curve and compare it with the peak in dashed curves in the figure \ref{fig9}(g)-(i) in the $\theta_{\mathbf{K}}$ region). For $\theta_{\rho}$ = 0 and $\frac{\pi}{4}$, we observe a destructive interference in the forward region (see figure \ref{fig9}(a)-(f) around $\theta_e$ = 0\textdegree region). On the contrary, for $\theta_{\rho}$ = $\frac{\pi}{2}$, we observe a constructive interference in the forward region for all {\it m} (see solid and dashed curves in figure \ref{fig9}(g)-(i)).

\subsection{Azimuthal angle dependence of $T_{fi}^{tw}$}

In figure \ref{fig6}, we studied the dependence of angular profiles of 5DCS on different values of {\it m} for different $\theta_{\rho}$. The study of the {\it T}-matrix element as a function of azimuthal angle, $\phi_p$ of the twisted electron beam is helpful for further exploration of the dependence of 5DCS on the OAM number {\it m}. From equation (\ref{e2}), it is clear that the 5DCS, is proportional to the square of the {\it T}-matrix element, $T_{fi}$. From figure \ref{fig6}(c) and \ref{fig6}(f), we observe that for $\theta_{\rho}$ = $\frac{\pi}{2}$ for both the kinematics, the angular profiles of the 5DCS for $m \neq$ 0 peaks around  $\theta_e$ = 187\textdegree. Also, we observe peak in the forward region ($\theta_e$ = 360\textdegree) for the kinematics $E_i$ = 4087 eV, $E_e$ = 20 eV, $\theta_s$ = 1\textdegree\ for $m \neq$ 0 (see figure \ref{fig6}(c)).  From figure \ref{fig6}, we observe that as \textit{m} increases from 1 to 3, the magnitude of the 5DCS decreases. The study of the angular profiles of the $T_{fi}^{tw}$ as a function of $\phi_p$ will be helpful for further investigation on this aspect. From equation (\ref{e24}), the transition matrix element $T^{tw}_{fi}(\varkappa,\mathbf{K})$, can also be expressed as;
\begin{equation*}
T^{tw}_{fi}(\varkappa,\mathbf{K}) = \int_0^{2\pi} d\phi_p T_{fi}^{tw}(\phi_p),
\end{equation*} 
with $T_{fi}^{tw}(\phi_p) \propto e^{im\phi_p} T_{fi}(\mathbf{K})$.

	Figure \ref{fig10}(a)-(b) and (c)-(d) represents the Re($T_{fi}^{tw}(\phi_p)$) and Im($T_{fi}^{tw}{\phi_p}$) as a function of $\phi_p$ for $\theta_e$ = 187\textdegree\ (backward peak) and $\theta_e$ = 360\textdegree\ (forward peak) for $\theta_{\rho}$ = $\frac{\pi}{2}$ for the kinematics $E_i$ = 4087 eV, $E_e$ = 20 eV, $\theta_s$ = 1\textdegree. While figure \ref{fig10}(e) and (f) represent the Re($T_{fi}^{tw}(\phi_p)$) and Im($T_{fi}^{tw}(\phi_p)$) for $\theta_e$ = 187\textdegree\ (backward region)  for $\theta_{\rho}$ = $\frac{\pi}{2}$ for the kinematics $E_i$ = 4168 eV, $E_e$ = 100 eV, $\theta_s$ = 8.2\textdegree. The dashed, dotted and dashed-dotted represents the contributions from {\it m} = 1, 2 and 3 respectively. From figure \ref{fig10}(a) and (b), we observe that the oscillations in both the real and imaginary part of $T_{fi}^{tw}$ are more pronounced for higher {\it m} and thus making less contributions to the matrix element $T_{fi}^{tw}$ when integrated over $\phi_p$ (see dashed, dotted and dashed-dotted curves in figure \ref{fig6}). We observe the similar trends for figure \ref{fig10}(c), (d) and (e), (f). 
	Since, $|T_{fi}^{tw}(\varkappa, \mathbf{K})|^2$ = $|\int_0^{2\pi} Re(T_{fi}^{tw}(\phi_p))d\phi_p|^2$ + $\int_0^{2\pi}|Im(T_{fi}^{tw}(\phi_p)) d\phi_p|^2$, the sum of area under the curves for $Re(T_{fi}^{tw}(\phi_p))$ and $Im(T_{fi}^{tw}(\phi_p))$ is an important quantity contributing to the 5DCS. In all the cases of figure (\ref{fig10}), we observe that due to increased oscillations, the area under the curve decreases as {\it m} increases. This results in decreased magnitude of 5DCS (5DCS $\propto  |T_{fi}^{tw}(\varkappa, \mathbf{K})|^2$) for higher {\it m}.

\subsection{$(TDCS)_{av}$ for macroscopic $H_2$ molecular target} 

In figure \ref{fig11}, we present the results for our calculations for the TDCS averaged over the impact parameter $\mathbf{b}$. From equation (\ref{e24}) and (\ref{e2}), we can compute the TDCS for an (e,2e) process on $H_2$ molecule for specific $\mathbf{b}$. The $(TDCS)_{av}$, averaged over the impact parameters, mimic more realistic scenario for the experiments with the twisted electron on the macroscopic target. As explained in the equation (\ref{e29}), the $(TDCS)_{av}$ for such case depends on the opening angle $\theta_p$ of the incident twisted electron beam (see equation (\ref{e29})). Figure \ref{fig11}(a) represents the $(TDCS)_{av}$ for $E_i$ = 4087 eV, $E_e$ = 20 eV and $\theta_s$ = 1\textdegree\ while figure \ref{fig11}(b) is for the kinematics $E_i$ = 4168 eV, $E_e$ = 100 eV and $\theta_s$ = 8.2\textdegree. We compare these results with the plane wave $(TDCS)_{av}$ case. In figure \ref{fig11}(a), we represent the angular profiles of the $(TDCS)_{av}$ for  different opening angles $\theta_p$, {\it viz.} 0.5\textdegree, 1\textdegree, 2\textdegree and 4\textdegree\ by dashed, dotted, dashed-dotted and dashed-dotted-dotted curves respectively.  We have scaled up the magnitude of the $(TDCS)_{av}$ by  a factor of 3 and 5 for $\theta_p$ = 2\textdegree\ and 4\textdegree\ to compare them with the plane wave results. We observed that for smaller value of $\theta_p$, like 0.5\textdegree, the angular profile of the $(TDCS)_{av}$ is similar to that for the plane wave (see solid and dashed curve in figure \ref{fig11}(a)).  However, for $\theta_p$ = 1\textdegree (also here $\theta_p$ = $\theta_s$ = 1\textdegree), the angular profile of the $(TDCS)_{av}$ is different from that of the plane wave (see dotted curve in figure \ref{fig11}(a)). In this case, we have obtained peaks in the forward and backward regions. With a further increase in the opening angle ($\theta_p$), we observe peaks in the perpendicular directions (see dashed-dotted and dashed-dotted-dotted curves in figure \ref{fig11}(a) around $\theta_e$ = 90\textdegree\ and 270\textdegree) with profound shift from the plane wave peaks. When we increase the opening angle ($\theta_p$) of the incident twisted beam, larger momentum is transferred to the target in the perpendicular direction which results in the formation of peaks in the perpendicular direction.  Further, when we increase $\theta_p$ the $(TDCS)_{av}$ decreases since we scale up the $(TDCS)_{av}$ for $\theta_p$ = 2\textdegree\ and 4\textdegree\ by a factor of 3 and 5 respectively.

 In figure \ref{fig11}(b), we discuss the $(TDCS)_{av}$ for the kinematics $E_i$ = 4168 eV, $E_e$ = 100 eV and $\theta_s$ = 8.2\textdegree\ for different values of $\theta_p$ as shown on the figure. Here also, we observe that for smaller $\theta_p$, the angular profile of the $(TDCS)_{av}$ is quite similar to that of the plane wave (see solid and dashed curve in figure \ref{fig11}(b)). For this kinematics also, when we increase the opening angle ($\theta_p$) of the twisted beam, the magnitude of the $(TDCS)_{av}$ decreases as we have scaled up the $(TDCS)_{av}$ by a factor of 10, 500 and 2000 for $\theta_p$ = 4\textdegree, 8.2\textdegree and 10\textdegree\ respectively (see dotted, dashed-dotted and dashed-dotted-dotted curves in figure \ref{fig11}(b)) to compare them with the plane wave results. For all the cases of $\theta_p$, except for $\theta_p$ = $\theta_s$ = 8.2\textdegree, we obtain peaks in the binary peak region. For the larger $\theta_p$, the binary peaks are shifted in perpendicular direction (towards $\theta_e$ = 270\textdegree). Like in the previous case, for $\theta_p$ = 8.2\textdegree, we also observe a dominant peak in the forward region and small peak in the backward region (see dashed-dotted curve in the figure \ref{fig11}(b)). In both the cases, we observe a dip in the $(TDCS)_{av}$ in perpendicular direction for $\theta_p$ = $\theta_s$ (see dotted curve in figure \ref{fig11}(a) and dashed-dotted curve in figure \ref{fig11}(b) around $\theta_e$ = 90\textdegree).


\section{Conclusion}\label{sec4}
In this paper, we have presented the theoretical study of five-fold differential cross sections(5DCS) and triple differential cross sections(TDCS) for an (e,2e) process on $H_2$ molecule by the twisted electron beam. We studied the angular distributions of the  5DCS and TDCS for the coplanar asymmetric geometry in the first Born approximation for both the plane and twisted electron wave. We have studied the effect of the OAM number, {\it m}, on the 5DCS and TDCS for {\it m} = 1, 2 and 3. We have benchmarked our theoretical results with the experimental data for the plane wave electron beam. We observed the dependence of the angular profiles of the TDCS on {\it m} for both small and large momentum transfer. The magnitude of the 5DCS, for three different orientations of the internuclear vector $\boldsymbol{\rho}$, drops significantly for higher {\it m}. We observed that the interference patterns in the angular profiles of the 5DCS depend on both the OAM number {\it m} and the orientation of the $H_2$ target. We discuss the dependence of the {\it T}-matrix element, $T_{fi}^{tw}(\phi_p)$ on the azimuthal angle ($\phi_p$) of the twisted electron beam. We observed that for a higher {\it m} there are pronounced oscillations in real and imaginary parts of the $T_{fi}(\phi_p)$ resulting in decreased 5DCS. We also discuss the $(TDCS)_{av}$ (averaged over the impact parameter \textbf{b}), of $H_2$ as a function of the opening angle, $\theta_p$ of the twisted electron beam. We observe that the angular profile of $(TDCS)_{av}$ significantly depends on the opening angle ($\theta_p$) of the twisted electron beam.

	Our present communication is the first attempt for an (e,2e) process on the $H_2$ molecule to unfold the effects of the twisted electron's different parameters on the angular profile of 5DCS and TDCS. The present study can also be extended for other complicated molecular targets, like $N_2$, $O_2$, $CH_4$ etc.  Besides this,  one can further explore  a theoretical model which is beyond the first Born approximation for the twisted electron (e,2e) process. One can also explore the differential cross-sections of the $H_2$ molecule for the twisted electrons for more sophisticated models, like TEC, M3DW, MBBK, and MCM3DW, etc. It is worth mentioning that we do not have experimental data of the (e,2e) process on $H_2$ molecule for the twisted electron case to confirm our results. We are hopeful that in the near future, the present study may prompt theoretical and experimental studies for electron impact ionization with twisted electron beams.

\section*{Acknowledgments}
Authors acknowledge the Science and Engineering Research Board, Department of Science and Technology, Government of India, for funding the project through the Grant number EMR/2016/003125.


\begin{thebibliography}{10}
\expandafter\ifx\csname url\endcsname\relax
  \def\url#1{{\tt #1}}\fi
\expandafter\ifx\csname urlprefix\endcsname\relax\def\urlprefix{ }\fi
\providecommand{\eprint}[2][]{\url{#2}}

\bibitem{Bart2016}
Bartschat K and Kushner M~J 2016 {\em Proc. Natl. Acad. Sci. U.S.A\/}
  \href{http://dx.doi.org/https://doi.org/10.1073/pnas.1606132113}{{\bf 113}
  7026--7034}

\bibitem{Shalenov2017}
Shalenov E, Dzhumagulova K and Ramazanov T 2017 {\em Phys. Plasmas\/}
  \href{http://dx.doi.org/https://doi.org/10.1063/1.4973324}{{\bf 24} 012101}

\bibitem{Bug2017}
Bug M~U, Baek W~Y, Rabus H, Villagrasa C, Meylan S and Rosenfeld A~B 2017 {\em
  Radiat. Phys. Chem.\/}
  \href{http://dx.doi.org/https://doi.org/10.1016/j.radphyschem.2016.09.027}{{\bf
  130} 459--479}

\bibitem{De2019}
de~Avillez M~A, Guerra M, Santos J~P and Breitschwerdt D 2019 {\em Astron.
  Astrophys.\/}
  \href{http://dx.doi.org/https://doi.org/10.1051/0004-6361/201935337}{{\bf
  631} A42}

\bibitem{Chavez2019}
Chavez A, Gimeno O, Rey A, Pliego G, Oropesa A, Alvarez P and Beltran F 2019
  {\em Chem. Eng. J.\/}
  \href{http://dx.doi.org/https://doi.org/10.1016/j.cej.2018.12.064}{{\bf 361}
  89--98}

\bibitem{Camp2018}
Campeanu R~I, Walters H~R~J and Whelan C~T 2018 {\em Phys. Rev. A\/}
  \href{http://dx.doi.org/10.1103/PhysRevA.97.062702}{{\bf 97} 062702}

\bibitem{Amaldi1969}
Amaldi~Jr U, Egidi A, Marconero R and Pizzella G 1969 {\em Rev. Sci. Instrum\/}
  \href{http://dx.doi.org/https://doi.org/10.1063/1.1684135}{{\bf 40}
  1001--1004}

\bibitem{Ehrhardt1969}
Ehrhardt H, Schulz M, Tekaat T and Willmann K 1969 {\em Phys. Rev. Lett.\/}
  \href{http://dx.doi.org/10.1103/PhysRevLett.22.89}{{\bf 22} 89--92}

\bibitem{Ren2015}
Ren X, Senftleben A, Pfl\"uger T, Bartschat K, Zatsarinny O, Berakdar J, Colgan
  J, Pindzola M~S, Bray I, Fursa D~V and Dorn A 2015 {\em Phys. Rev. A\/}
  \href{http://dx.doi.org/10.1103/PhysRevA.92.052707}{{\bf 92} 052707}

\bibitem{Whelan2012}
Whelan C~T, Walters H, Lahmam-Bennani A and Ehrhardt H 2012 {\em (e, 2e) \&
  related processes\/} vol 414 (Springer Science \& Business Media)

\bibitem{Lahmam1991}
Lahmam-Bennani A 1991 {\em J. Phys. B: At. Mol. Opt. Phys.\/}
  \href{http://dx.doi.org/10.1088/0953-4075/24/10/001}{{\bf 24} 2401--2442}

\bibitem{Casagrande2008}
Casagrande E~M~S {\em et~al.\/} 2008 {\em J. Phys. Conf. Ser\/}
  \href{http://dx.doi.org/10.1088/1742-6596/141/1/012016}{{\bf 141} 012016}

\bibitem{Amami2017}
Amami S~M~F 2017 {\em Theoretical calculations for electron impact ionization
  of atoms and molecules\/} Ph.D. thesis Missouri University of Science and
  Technology,Missouri
  \urlprefix\url{https://scholarsmine.mst.edu/doctoral_dissertations/2617}

\bibitem{Colyer2009}
Colyer C, Stevenson M, Al-Hagan O, Madison D, Ning C and Lohmann B 2009 {\em J.
  Phys. B: At. Mol. Opt. Phys.\/}
  \href{http://dx.doi.org/https://doi.org/10.1088/0953-4075/42/23/235207}{{\bf
  42} 235207}

\bibitem{Colgan2009}
Colgan J, Al-Hagan O, Madison D, Murray A~J and Pindzola M 2009 {\em J. Phys.
  B: At. Mol. Opt. Phys.\/}
  \href{http://dx.doi.org/https://doi.org/10.1088/0953-4075/42/17/171001}{{\bf
  42} 171001}

\bibitem{Mouawad2017}
Mouawad L, Hervieux P~A, Cappello C~D, Pansanel J, Osman A, Khalil M and Bitar
  Z~E 2017 {\em J. Phys. B: At. Mol. Opt. Phys.\/}
  \href{http://dx.doi.org/10.1088/1361-6455/aa8cb9}{{\bf 50} 215204}

\bibitem{Ren2017}
Ren X, Amami S, Hossen K, Ali E, Ning C, Colgan J, Madison D and Dorn A 2017
  {\em Phys. Rev. A\/} \href{http://dx.doi.org/10.1103/PhysRevA.95.022701}{{\bf
  95} 022701}

\bibitem{Li2017}
Li X, Gong M, Liu L, Wu Y, Wang J, Qu Y and Chen X 2017 {\em Phys. Rev. A\/}
  \href{http://dx.doi.org/10.1103/PhysRevA.95.012703}{{\bf 95} 012703}

\bibitem{Hossen2018}
Hossen K, Ren X, Wang E, Gong M, Li X, Zhang S~B, Chen X and Dorn A 2018 {\em
  J. Phys. B: At. Mol. Opt. Phys.\/}
  \href{http://dx.doi.org/https://doi.org/10.1088/1361-6455/aae0ab}{{\bf 51}
  215201}

\bibitem{Sakaamini2018}
Sakaamini A, Harvey M, Amami S, Murray A~J, Madison D and Ning C 2018 {\em J.
  Phys. B: At. Mol. Opt. Phys.\/}
  \href{http://dx.doi.org/https://doi.org/10.1088/1361-6455/aa9eb9}{{\bf 51}
  035207}

\bibitem{Khatir2019}
Khatir T, Houamer S and Dal~Cappello C 2019 {\em J. Phys. B: At. Mol. Opt.
  Phys.\/}
  \href{http://dx.doi.org/https://doi.org/10.1088/1361-6455/ab4a70}{{\bf 52}
  245201}

\bibitem{Singh2019}
Singh P, Purohit G, Champion C, S{\'e}billeau D and Madison D 2019 {\em J.
  Chem. Phys.\/}
  \href{http://dx.doi.org/https://doi.org/10.1063/1.5088966}{{\bf 150} 054304}

\bibitem{Ali2020}
Ali E, Chakraborty H and Madison D 2020 {\em J. Chem. Phys.\/}
  \href{http://dx.doi.org/https://doi.org/10.1063/1.5143148}{{\bf 152} 124303}

\bibitem{Mansouri2004}
Mansouri A, Cappello C~D, Houamer S, Charpentier I and Lahmam-Bennani A 2004
  {\em J. Phys. B: At. Mol. Opt. Phys.\/}
  \href{http://dx.doi.org/10.1088/0953-4075/37/6/006}{{\bf 37} 1203--1214}

\bibitem{Chul2011}
Chuluunbaatar O, Gusev A and Joulakian B 2011 {\em J. Phys. B: At. Mol. Opt.
  Phys.\/}
  \href{http://dx.doi.org/https://doi.org/10.1088/0953-4075/45/1/015205}{{\bf
  45} 015205}

\bibitem{Pindzola2018}
Pindzola M~S, Colgan J~P and McLaughlin B~M 2018 {\em J. Phys. B: At. Mol. Opt.
  Phys.\/}
  \href{http://dx.doi.org/https://doi.org/10.1088/1361-6455/aaa2d2}{{\bf 51}
  035206}

\bibitem{Cohen1966}
Cohen H~D and Fano U 1966 {\em Phys. Rev.\/}
  \href{http://dx.doi.org/10.1103/PhysRev.150.30}{{\bf 150} 30--33}

\bibitem{Stolterfoht2001}
Stolterfoht N {\em et~al.\/} 2001 {\em Phys. Rev. Lett.\/}
  \href{http://dx.doi.org/10.1103/PhysRevLett.87.023201}{{\bf 87} 023201}

\bibitem{Stolterfoht2003}
Stolterfoht N {\em et~al.\/} 2003 {\em Phys. Rev. A\/}
  \href{http://dx.doi.org/10.1103/PhysRevA.67.030702}{{\bf 67} 030702}

\bibitem{Brownlie2006}
Milne-Brownlie D~S, Foster M, Gao J, Lohmann B and Madison D~H 2006 {\em Phys.
  Rev. Lett.\/} \href{http://dx.doi.org/10.1103/PhysRevLett.96.233201}{{\bf 96}
  233201}

\bibitem{Staicu2008}
Casagrande E~M~S {\em et~al.\/} 2008 {\em J. Phys. B: At. Mol. Opt. Phys.\/}
  \href{http://dx.doi.org/10.1088/0953-4075/41/2/025204}{{\bf 41} 025204}

\bibitem{Fojon2006}
Foj{\'o}n O, Stia C and Rivarola R 2006 {\em AIP Conf. Proc.\/}
  \href{http://dx.doi.org/https://doi.org/10.1063/1.2165618}{{\bf 811} 42--47}

\bibitem{Stia2003}
Stia C, Foj{\'o}n O, Weck P, Hanssen J and Rivarola R 2003 {\em J. Phys. B: At.
  Mol. Opt. Phys.\/} \href{http://dx.doi.org/10.1088/0953-4075/36/17/101}{{\bf
  36} L257}

\bibitem{Ciappina2014}
Ciappina M~F, Foj{\'{o}}n O~A and Rivarola R~D 2014 {\em J. Phys. B: At. Mol.
  Opt. Phys.\/} \href{http://dx.doi.org/10.1088/0953-4075/47/4/042001}{{\bf 47}
  042001} \urlprefix\url{https://doi.org/10.1088%2F0953-4075%2F47%2F4%2F042001}

\bibitem{Li2018}
Li X, Ren X, Hossen K, Wang E, Chen X and Dorn A 2018 {\em Phys. Rev. A\/}
  \href{http://dx.doi.org/10.1103/PhysRevA.97.022706}{{\bf 97} 022706}

\bibitem{Bliokh2007}
Bliokh K~Y, Bliokh Y~P, Savel'ev S and Nori F 2007 {\em Phys. Rev. Lett.\/}
  \href{http://dx.doi.org/10.1103/PhysRevLett.99.190404}{{\bf 99} 190404}

\bibitem{Uchida2010}
Uchida M and Tonomura A 2010 {\em Nature\/}
  \href{http://dx.doi.org/https://doi.org/10.1038/nature08904}{{\bf 464} 737}

\bibitem{Verbeeck2010}
Verbeeck J, Tian H and Schattschneider P 2010 {\em Nature\/}
  \href{http://dx.doi.org/https://doi.org/10.1038/nature09366}{{\bf 467} 301}

\bibitem{Morran2011}
McMorran B~J {\em et~al.\/} 2011 {\em Science\/}
  \href{http://dx.doi.org/10.1126/science.1198804}{{\bf 331} 192--195}

\bibitem{Llyod2017}
Lloyd S~M, Babiker M, Thirunavukkarasu G and Yuan J 2017 {\em Rev. Mod.
  Phys.\/} \href{http://dx.doi.org/10.1103/RevModPhys.89.035004}{{\bf 89}
  035004}

\bibitem{Bliokh2017}
Bliokh K~Y, Ivanov I~P, Guzzinati G, Clark L, Van~Boxem R, B{\'e}ch{\'e} A,
  Juchtmans R, Alonso M~A, Schattschneider P, Nori F {\em et~al.\/} 2017 {\em
  Physics Reports\/}
  \href{http://dx.doi.org/https://doi.org/10.1016/j.physrep.2017.05.006}{{\bf
  690} 1--70}

\bibitem{Hugo2018}
Larocque H, Kaminer I, Grillo V, Leuchs G, Padgett M~J, Boyd R~W, Segev M and
  Karimi E 2018 {\em Contemp. Phys\/}
  \href{http://dx.doi.org/10.1080/00107514.2017.1418046}{{\bf 59} 126--144}

\bibitem{Neil2002}
O'Neil A~T, MacVicar I, Allen L and Padgett M~J 2002 {\em Phys. Rev. Lett.\/}
  \href{http://dx.doi.org/10.1103/PhysRevLett.88.053601}{{\bf 88} 053601}

\bibitem{Spiral2005}
Furhapter S, Jesacher A, Bernet S and Ritsch-Marte M 2005 {\em Opt. Express\/}
  \href{http://dx.doi.org/https://doi.org/10.1364/OPEX.13.000689}{{\bf 13}
  689--694}

\bibitem{Berkhout2009}
Berkhout G and Beijersbergen M 2009 {\em J. Opt\/}
  \href{http://dx.doi.org/https://doi.org/10.1088/1464-4258/11/9/094021}{{\bf
  11} 094021}

\bibitem{Sebastian2019}
Gemsheim S and Rost J~M 2019 {\em Phys. Rev. A\/}
  \href{http://dx.doi.org/10.1103/PhysRevA.100.043408}{{\bf 100} 043408}

\bibitem{Serbo2011}
Ivanov I~P and Serbo V~G 2011 {\em Phys. Rev. A\/}
  \href{http://dx.doi.org/10.1103/PhysRevA.84.033804}{{\bf 84} 033804}

\bibitem{Boxem2014}
Van~Boxem R, Partoens B and Verbeeck J 2014 {\em Phys. Rev. A\/}
  \href{http://dx.doi.org/10.1103/PhysRevA.89.032715}{{\bf 89} 032715}

\bibitem{Boxem2015}
Van~Boxem R, Partoens B and Verbeeck J 2015 {\em Phys. Rev. A\/}
  \href{http://dx.doi.org/10.1103/PhysRevA.91.032703}{{\bf 91} 032703}

\bibitem{Serbo2015}
Serbo V, Ivanov I~P, Fritzsche S, Seipt D and Surzhykov A 2015 {\em Phys. Rev.
  A\/} \href{http://dx.doi.org/10.1103/PhysRevA.92.012705}{{\bf 92} 012705}

\bibitem{Karlovets2017}
Karlovets D~V, Kotkin G~L, Serbo V~G and Surzhykov A 2017 {\em Phys. Rev. A\/}
  \href{http://dx.doi.org/10.1103/PhysRevA.95.032703}{{\bf 95} 032703}

\bibitem{Maiorova2018}
Maiorova A~V, Fritzsche S, M\"uller R~A and Surzhykov A 2018 {\em Phys. Rev.
  A\/} \href{http://dx.doi.org/10.1103/PhysRevA.98.042701}{{\bf 98} 042701}

\bibitem{Harris2019}
Harris A, Plumadore A and Smozhanyk Z 2019 {\em J. Phys. B: At. Mol. Opt.
  Phys.\/}
  \href{http://dx.doi.org/https://doi.org/10.1088/1361-6455/ab12f3}{{\bf 52}
  094001}

\bibitem{Mandal2020}
Mandal A, Dhankhar N, S\'ebilleau D and Choubisa R 2020 (\textit{Preprint}
  \eprint{arXiv:2003.06459})

\bibitem{Dhankhar2020}
Dhankhar N, Mandal A and Choubisa R 2020 {\em J. Phys. B: At. Mol. Opt.
  Phys.\/} \href{http://dx.doi.org/10.1088/1361-6455/ab8718}{{\bf 53} 155203}

\bibitem{Wang1928}
Wang S~C 1928 {\em Phys. Rev.\/}
  \href{http://dx.doi.org/10.1103/PhysRev.31.579}{{\bf 31} 579--586}

\bibitem{Stia2002}
Stia C~R, Foj\'on O~A, Weck P~F, Hanssen J, Joulakian B and Rivarola R~D 2002
  {\em Phys. Rev. A\/} \href{http://dx.doi.org/10.1103/PhysRevA.66.052709}{{\bf
  66} 052709}

\bibitem{Tweed1992}
Tweed R 1992 {\em Z. Phys. D\/}
  \href{http://dx.doi.org/https://doi.org/10.1007/BF01429252}{{\bf 23}
  309--320}

\bibitem{Cherid1989}
Ch{\'e}rid M, Lahmam-Bennani A {\em et~al.\/} 1989 {\em J. Phys. B: At. Mol.
  Opt. Phys.\/}
  \href{http://dx.doi.org/https://doi.org/10.1088/0953-4075/22/21/012}{{\bf 22}
  3483}

\bibitem{Weck2001}
Weck P, Foj\'on O~A, Hanssen J, Joulakian B and Rivarola R~D 2001 {\em Phys.
  Rev. A\/} \href{http://dx.doi.org/10.1103/PhysRevA.63.042709}{{\bf 63}
  042709}

\bibitem{Gao2006}
Gao J, Madison D~H and Peacher J~L 2006 {\em J. Phys. B: At. Mol. Opt. Phys.\/}
  \href{http://dx.doi.org/10.1088/0953-4075/39/6/002}{{\bf 39} 1275--1284}

\bibitem{Ali2019}
Ali E and Madison D 2019 {\em Phys. Rev. A\/}
  \href{http://dx.doi.org/10.1103/PhysRevA.100.012712}{{\bf 100} 012712}

\end{thebibliography}
\providecommand{\newblock}{}

\end{document}